\documentclass[sigconf]{acmart}
\AtBeginDocument{%
  }

\usepackage{xcolor}  
\usepackage{algorithm}
\usepackage{algpseudocode}
\usepackage{setspace}
\usepackage{tabularx}

\newcommand{\h}[1]{{\it #1}}

\renewcommand\footnotetextcopyrightpermission[1]{}

\setcopyright{none}
\acmConference{arXiv preprint}{2025}{}
\acmYear{2025}

\begin{document}

\newcommand{\HEQwenIIIavgprobUncalibECE}{ 0.316}
\newcommand{\HEQwenIIIavgprobUncalibASCE}{ 0.104}
\newcommand{\HEQwenIIIavgprobUncalibMSE}{ 0.333}
\newcommand{\HEQwenIIIavgprobUncalibBrierref}{ 0.246}
\newcommand{\HEQwenIIIavgprobUncalibSkillScore}{ -0.350}
\newcommand{\HEQwenIIIavgprobUncalibACC}{ 0.568}
\newcommand{\HEQwenIIIavgprobPLATTECE}{ 0.036}
\newcommand{\HEQwenIIIavgprobPLATTASCE}{ 0.002}
\newcommand{\HEQwenIIIavgprobPLATTMSE}{ 0.231}
\newcommand{\HEQwenIIIavgprobPLATTBrierref}{ 0.246}
\newcommand{\HEQwenIIIavgprobPLATTSkillScore}{ 0.064}
\newcommand{\HEQwenIIIavgprobPLATTACC}{ 0.616}
\newcommand{\HEQwenIIIavgprobHBECE}{ 0.022}
\newcommand{\HEQwenIIIavgprobHBASCE}{ 0.001}
\newcommand{\HEQwenIIIavgprobHBMSE}{ 0.230}
\newcommand{\HEQwenIIIavgprobHBBrierref}{ 0.246}
\newcommand{\HEQwenIIIavgprobHBSkillScore}{ 0.066}
\newcommand{\HEQwenIIIavgprobHBACC}{ 0.612}
\newcommand{\HEQwenIIIavgprobLRECE}{ 0.035}
\newcommand{\HEQwenIIIavgprobLRASCE}{ 0.002}
\newcommand{\HEQwenIIIavgprobLRMSE}{ 0.172}
\newcommand{\HEQwenIIIavgprobLRBrierref}{ 0.246}
\newcommand{\HEQwenIIIavgprobLRSkillScore}{ 0.304}
\newcommand{\HEQwenIIIavgprobLRACC}{ 0.745}
\newcommand{\HEQwenIIIavgprobLOGRECE}{ 0.256}
\newcommand{\HEQwenIIIavgprobLOGRASCE}{ 0.066}
\newcommand{\HEQwenIIIavgprobLOGRMSE}{ 0.256}
\newcommand{\HEQwenIIIavgprobLOGRBrierref}{ 0.246}
\newcommand{\HEQwenIIIavgprobLOGRSkillScore}{ -0.039}
\newcommand{\HEQwenIIIavgprobLOGRACC}{ 0.744}
\newcommand{\HEQwenIIIavgprobIGHBECE}{ 0.244}
\newcommand{\HEQwenIIIavgprobIGHBASCE}{ 0.076}
\newcommand{\HEQwenIIIavgprobIGHBMSE}{ 0.305}
\newcommand{\HEQwenIIIavgprobIGHBBrierref}{ 0.246}
\newcommand{\HEQwenIIIavgprobIGHBSkillScore}{ -0.237}
\newcommand{\HEQwenIIIavgprobIGHBACC}{ 0.571}
\newcommand{\HEQwenIIIavgprobIGLBECE}{ 0.041}
\newcommand{\HEQwenIIIavgprobIGLBASCE}{ 0.003}
\newcommand{\HEQwenIIIavgprobIGLBMSE}{ 0.192}
\newcommand{\HEQwenIIIavgprobIGLBBrierref}{ 0.246}
\newcommand{\HEQwenIIIavgprobIGLBSkillScore}{ 0.221}
\newcommand{\HEQwenIIIavgprobIGLBACC}{ 0.715}

\newcommand{\HEGPTOSSavgprobUncalibECE}{ 0.275}
\newcommand{\HEGPTOSSavgprobUncalibASCE}{ 0.096}
\newcommand{\HEGPTOSSavgprobUncalibMSE}{ 0.136}
\newcommand{\HEGPTOSSavgprobUncalibBrierref}{ 0.044}
\newcommand{\HEGPTOSSavgprobUncalibSkillScore}{ -2.073}
\newcommand{\HEGPTOSSavgprobUncalibACC}{ 0.831}
\newcommand{\HEGPTOSSavgprobPLATTECE}{ 0.021}
\newcommand{\HEGPTOSSavgprobPLATTASCE}{ 0.001}
\newcommand{\HEGPTOSSavgprobPLATTMSE}{ 0.041}
\newcommand{\HEGPTOSSavgprobPLATTBrierref}{ 0.044}
\newcommand{\HEGPTOSSavgprobPLATTSkillScore}{ 0.069}
\newcommand{\HEGPTOSSavgprobPLATTACC}{ 0.954}
\newcommand{\HEGPTOSSavgprobHBECE}{ 0.020}
\newcommand{\HEGPTOSSavgprobHBASCE}{ 0.001}
\newcommand{\HEGPTOSSavgprobHBMSE}{ 0.041}
\newcommand{\HEGPTOSSavgprobHBBrierref}{ 0.044}
\newcommand{\HEGPTOSSavgprobHBSkillScore}{ 0.069}
\newcommand{\HEGPTOSSavgprobHBACC}{ 0.954}
\newcommand{\HEGPTOSSavgprobLRECE}{ 0.038}
\newcommand{\HEGPTOSSavgprobLRASCE}{ 0.003}
\newcommand{\HEGPTOSSavgprobLRMSE}{ 0.039}
\newcommand{\HEGPTOSSavgprobLRBrierref}{ 0.044}
\newcommand{\HEGPTOSSavgprobLRSkillScore}{ 0.115}
\newcommand{\HEGPTOSSavgprobLRACC}{ 0.954}
\newcommand{\HEGPTOSSavgprobLOGRECE}{ 0.045}
\newcommand{\HEGPTOSSavgprobLOGRASCE}{ 0.004}
\newcommand{\HEGPTOSSavgprobLOGRMSE}{ 0.045}
\newcommand{\HEGPTOSSavgprobLOGRBrierref}{ 0.044}
\newcommand{\HEGPTOSSavgprobLOGRSkillScore}{ -0.024}
\newcommand{\HEGPTOSSavgprobLOGRACC}{ 0.955}
\newcommand{\HEGPTOSSavgprobIGHBECE}{ 0.253}
\newcommand{\HEGPTOSSavgprobIGHBASCE}{ 0.086}
\newcommand{\HEGPTOSSavgprobIGHBMSE}{ 0.127}
\newcommand{\HEGPTOSSavgprobIGHBBrierref}{ 0.044}
\newcommand{\HEGPTOSSavgprobIGHBSkillScore}{ -1.856}
\newcommand{\HEGPTOSSavgprobIGHBACC}{ 0.831}
\newcommand{\HEGPTOSSavgprobIGLBECE}{ 0.035}
\newcommand{\HEGPTOSSavgprobIGLBASCE}{ 0.003}
\newcommand{\HEGPTOSSavgprobIGLBMSE}{ 0.040}
\newcommand{\HEGPTOSSavgprobIGLBBrierref}{ 0.044}
\newcommand{\HEGPTOSSavgprobIGLBSkillScore}{ 0.103}
\newcommand{\HEGPTOSSavgprobIGLBACC}{ 0.952}

\newcommand{\HEDSRavgprobUncalibECE}{ 0.442}
\newcommand{\HEDSRavgprobUncalibASCE}{ 0.210}
\newcommand{\HEDSRavgprobUncalibMSE}{ 0.314}
\newcommand{\HEDSRavgprobUncalibBrierref}{ 0.119}
\newcommand{\HEDSRavgprobUncalibSkillScore}{ -1.634}
\newcommand{\HEDSRavgprobUncalibACC}{ 0.414}
\newcommand{\HEDSRavgprobPLATTECE}{ 0.020}
\newcommand{\HEDSRavgprobPLATTASCE}{ 0.001}
\newcommand{\HEDSRavgprobPLATTMSE}{ 0.104}
\newcommand{\HEDSRavgprobPLATTBrierref}{ 0.119}
\newcommand{\HEDSRavgprobPLATTSkillScore}{ 0.125}
\newcommand{\HEDSRavgprobPLATTACC}{ 0.861}
\newcommand{\HEDSRavgprobHBECE}{ 0.011}
\newcommand{\HEDSRavgprobHBASCE}{ 0.001}
\newcommand{\HEDSRavgprobHBMSE}{ 0.105}
\newcommand{\HEDSRavgprobHBBrierref}{ 0.119}
\newcommand{\HEDSRavgprobHBSkillScore}{ 0.119}
\newcommand{\HEDSRavgprobHBACC}{ 0.861}
\newcommand{\HEDSRavgprobLRECE}{ 0.040}
\newcommand{\HEDSRavgprobLRASCE}{ 0.003}
\newcommand{\HEDSRavgprobLRMSE}{ 0.090}
\newcommand{\HEDSRavgprobLRBrierref}{ 0.119}
\newcommand{\HEDSRavgprobLRSkillScore}{ 0.242}
\newcommand{\HEDSRavgprobLRACC}{ 0.874}
\newcommand{\HEDSRavgprobLOGRECE}{ 0.120}
\newcommand{\HEDSRavgprobLOGRASCE}{ 0.020}
\newcommand{\HEDSRavgprobLOGRMSE}{ 0.120}
\newcommand{\HEDSRavgprobLOGRBrierref}{ 0.119}
\newcommand{\HEDSRavgprobLOGRSkillScore}{ -0.010}
\newcommand{\HEDSRavgprobLOGRACC}{ 0.880}
\newcommand{\HEDSRavgprobIGHBECE}{ 0.246}
\newcommand{\HEDSRavgprobIGHBASCE}{ 0.104}
\newcommand{\HEDSRavgprobIGHBMSE}{ 0.208}
\newcommand{\HEDSRavgprobIGHBBrierref}{ 0.119}
\newcommand{\HEDSRavgprobIGHBSkillScore}{ -0.747}
\newcommand{\HEDSRavgprobIGHBACC}{ 0.702}
\newcommand{\HEDSRavgprobIGLBECE}{ 0.043}
\newcommand{\HEDSRavgprobIGLBASCE}{ 0.004}
\newcommand{\HEDSRavgprobIGLBMSE}{ 0.101}
\newcommand{\HEDSRavgprobIGLBBrierref}{ 0.119}
\newcommand{\HEDSRavgprobIGLBSkillScore}{ 0.151}
\newcommand{\HEDSRavgprobIGLBACC}{ 0.870}

\newcommand{\LCBQwenIIIcodeprobUncalibECE}{0.342}
\newcommand{\LCBQwenIIIcodeprobUncalibASCE}{0.135}
\newcommand{\LCBQwenIIIcodeprobUncalibMSE}{0.337}
\newcommand{\LCBQwenIIIcodeprobUncalibBrierref}{0.232}
\newcommand{\LCBQwenIIIcodeprobUncalibSkillScore}{-0.453}
\newcommand{\LCBQwenIIIcodeprobUncalibACC}{0.634}
\newcommand{\LCBQwenIIIcodeprobHBECE}{0.014}
\newcommand{\LCBQwenIIIcodeprobHBASCE}{0.001}
\newcommand{\LCBQwenIIIcodeprobHBMSE}{0.204}
\newcommand{\LCBQwenIIIcodeprobHBBrierref}{0.232}
\newcommand{\LCBQwenIIIcodeprobHBSkillScore}{0.119}
\newcommand{\LCBQwenIIIcodeprobHBACC}{0.680}
\newcommand{\LCBQwenIIIcodeprobLRECE}{0.082}
\newcommand{\LCBQwenIIIcodeprobLRASCE}{0.009}
\newcommand{\LCBQwenIIIcodeprobLRMSE}{0.139}
\newcommand{\LCBQwenIIIcodeprobLRBrierref}{0.232}
\newcommand{\LCBQwenIIIcodeprobLRSkillScore}{0.402}
\newcommand{\LCBQwenIIIcodeprobLRACC}{0.817}
\newcommand{\LCBQwenIIIcodeprobLOGRECE}{0.194}
\newcommand{\LCBQwenIIIcodeprobLOGRASCE}{0.040}
\newcommand{\LCBQwenIIIcodeprobLOGRMSE}{0.194}
\newcommand{\LCBQwenIIIcodeprobLOGRBrierref}{0.232}
\newcommand{\LCBQwenIIIcodeprobLOGRSkillScore}{0.165}
\newcommand{\LCBQwenIIIcodeprobLOGRACC}{0.806}
\newcommand{\LCBQwenIIIcodeprobIGHBECE}{0.089}
\newcommand{\LCBQwenIIIcodeprobIGHBASCE}{0.024}
\newcommand{\LCBQwenIIIcodeprobIGHBMSE}{0.195}
\newcommand{\LCBQwenIIIcodeprobIGHBBrierref}{0.232}
\newcommand{\LCBQwenIIIcodeprobIGHBSkillScore}{0.161}
\newcommand{\LCBQwenIIIcodeprobIGHBACC}{0.698}
\newcommand{\LCBQwenIIIcodeprobIGLBECE}{0.093}
\newcommand{\LCBQwenIIIcodeprobIGLBASCE}{0.012}
\newcommand{\LCBQwenIIIcodeprobIGLBMSE}{0.145}
\newcommand{\LCBQwenIIIcodeprobIGLBBrierref}{0.232}
\newcommand{\LCBQwenIIIcodeprobIGLBSkillScore}{0.377}
\newcommand{\LCBQwenIIIcodeprobIGLBACC}{0.779}

\newcommand{\LCBQwenIIIavgprobUncalibECE}{0.429}
\newcommand{\LCBQwenIIIavgprobUncalibASCE}{0.233}
\newcommand{\LCBQwenIIIavgprobUncalibMSE}{0.383}
\newcommand{\LCBQwenIIIavgprobUncalibBrierref}{0.248}
\newcommand{\LCBQwenIIIavgprobUncalibSkillScore}{-0.543}
\newcommand{\LCBQwenIIIavgprobUncalibACC}{0.458}
\newcommand{\LCBQwenIIIavgprobUncalibgASCEcompeasy}{0.040}
\newcommand{\LCBQwenIIIavgprobUncalibgASCEcompmedium}{0.182}
\newcommand{\LCBQwenIIIavgprobUncalibgASCEcomphard}{0.578}
\newcommand{\LCBQwenIIIavgprobUncalibgASCEpromptlenhigh}{0.352}
\newcommand{\LCBQwenIIIavgprobUncalibgASCElochigh}{0.209}
\newcommand{\LCBQwenIIIavgprobUncalibgASCElenhigh}{0.424}
\newcommand{\LCBQwenIIIavgprobPLATTECE}{0.078}
\newcommand{\LCBQwenIIIavgprobPLATTASCE}{0.009}
\newcommand{\LCBQwenIIIavgprobPLATTMSE}{0.155}
\newcommand{\LCBQwenIIIavgprobPLATTBrierref}{0.248}
\newcommand{\LCBQwenIIIavgprobPLATTSkillScore}{0.377}
\newcommand{\LCBQwenIIIavgprobPLATTACC}{0.798}
\newcommand{\LCBQwenIIIavgprobPLATTgASCEcompeasy}{0.036}
\newcommand{\LCBQwenIIIavgprobPLATTgASCEcompmedium}{0.012}
\newcommand{\LCBQwenIIIavgprobPLATTgASCEcomphard}{0.056}
\newcommand{\LCBQwenIIIavgprobPLATTgASCEpromptlenhigh}{0.014}
\newcommand{\LCBQwenIIIavgprobPLATTgASCElochigh}{0.011}
\newcommand{\LCBQwenIIIavgprobPLATTgASCElenhigh}{0.012}
\newcommand{\LCBQwenIIIavgprobHBECE}{0.049}
\newcommand{\LCBQwenIIIavgprobHBASCE}{0.003}
\newcommand{\LCBQwenIIIavgprobHBMSE}{0.153}
\newcommand{\LCBQwenIIIavgprobHBBrierref}{0.248}
\newcommand{\LCBQwenIIIavgprobHBSkillScore}{0.383}
\newcommand{\LCBQwenIIIavgprobHBACC}{0.797}
\newcommand{\LCBQwenIIIavgprobHBgASCEcompeasy}{0.026}
\newcommand{\LCBQwenIIIavgprobHBgASCEcompmedium}{0.005}
\newcommand{\LCBQwenIIIavgprobHBgASCEcomphard}{0.059}
\newcommand{\LCBQwenIIIavgprobHBgASCEpromptlenhigh}{0.011}
\newcommand{\LCBQwenIIIavgprobHBgASCElochigh}{0.006}
\newcommand{\LCBQwenIIIavgprobHBgASCElenhigh}{0.011}
\newcommand{\LCBQwenIIIavgprobLRECE}{0.071}
\newcommand{\LCBQwenIIIavgprobLRASCE}{0.007}
\newcommand{\LCBQwenIIIavgprobLRMSE}{0.133}
\newcommand{\LCBQwenIIIavgprobLRBrierref}{0.248}
\newcommand{\LCBQwenIIIavgprobLRSkillScore}{0.463}
\newcommand{\LCBQwenIIIavgprobLRACC}{0.822}
\newcommand{\LCBQwenIIIavgprobLRgASCEcompeasy}{0.008}
\newcommand{\LCBQwenIIIavgprobLRgASCEcompmedium}{0.018}
\newcommand{\LCBQwenIIIavgprobLRgASCEcomphard}{0.014}
\newcommand{\LCBQwenIIIavgprobLRgASCEpromptlenhigh}{0.008}
\newcommand{\LCBQwenIIIavgprobLRgASCElochigh}{0.011}
\newcommand{\LCBQwenIIIavgprobLRgASCElenhigh}{0.008}
\newcommand{\LCBQwenIIIavgprobLOGRECE}{0.172}
\newcommand{\LCBQwenIIIavgprobLOGRASCE}{0.030}
\newcommand{\LCBQwenIIIavgprobLOGRMSE}{0.172}
\newcommand{\LCBQwenIIIavgprobLOGRBrierref}{0.248}
\newcommand{\LCBQwenIIIavgprobLOGRSkillScore}{0.306}
\newcommand{\LCBQwenIIIavgprobLOGRACC}{0.828}
\newcommand{\LCBQwenIIIavgprobLOGRgASCEcompeasy}{0.033}
\newcommand{\LCBQwenIIIavgprobLOGRgASCEcompmedium}{0.068}
\newcommand{\LCBQwenIIIavgprobLOGRgASCEcomphard}{0.004}
\newcommand{\LCBQwenIIIavgprobLOGRgASCEpromptlenhigh}{0.032}
\newcommand{\LCBQwenIIIavgprobLOGRgASCElochigh}{0.044}
\newcommand{\LCBQwenIIIavgprobLOGRgASCElenhigh}{0.021}
\newcommand{\LCBQwenIIIavgprobIGHBECE}{0.148}
\newcommand{\LCBQwenIIIavgprobIGHBASCE}{0.045}
\newcommand{\LCBQwenIIIavgprobIGHBMSE}{0.191}
\newcommand{\LCBQwenIIIavgprobIGHBBrierref}{0.248}
\newcommand{\LCBQwenIIIavgprobIGHBSkillScore}{0.232}
\newcommand{\LCBQwenIIIavgprobIGHBACC}{0.731}
\newcommand{\LCBQwenIIIavgprobIGHBgASCEcompeasy}{0.032}
\newcommand{\LCBQwenIIIavgprobIGHBgASCEcompmedium}{0.047}
\newcommand{\LCBQwenIIIavgprobIGHBgASCEcomphard}{0.131}
\newcommand{\LCBQwenIIIavgprobIGHBgASCEpromptlenhigh}{0.069}
\newcommand{\LCBQwenIIIavgprobIGHBgASCElochigh}{0.031}
\newcommand{\LCBQwenIIIavgprobIGHBgASCElenhigh}{0.058}
\newcommand{\LCBQwenIIIavgprobIGLBECE}{0.038}
\newcommand{\LCBQwenIIIavgprobIGLBASCE}{0.004}
\newcommand{\LCBQwenIIIavgprobIGLBMSE}{0.129}
\newcommand{\LCBQwenIIIavgprobIGLBBrierref}{0.248}
\newcommand{\LCBQwenIIIavgprobIGLBSkillScore}{0.480}
\newcommand{\LCBQwenIIIavgprobIGLBACC}{0.830}
\newcommand{\LCBQwenIIIavgprobIGLBgASCEcompeasy}{0.021}
\newcommand{\LCBQwenIIIavgprobIGLBgASCEcompmedium}{0.009}
\newcommand{\LCBQwenIIIavgprobIGLBgASCEcomphard}{0.006}
\newcommand{\LCBQwenIIIavgprobIGLBgASCEpromptlenhigh}{0.007}
\newcommand{\LCBQwenIIIavgprobIGLBgASCElochigh}{0.006}
\newcommand{\LCBQwenIIIavgprobIGLBgASCElenhigh}{0.010}

\newcommand{\LCBfQwenIIIavgprobUncalibECE}{0.294}
\newcommand{\LCBfQwenIIIavgprobUncalibASCE}{0.127}
\newcommand{\LCBfQwenIIIavgprobUncalibMSE}{0.292}
\newcommand{\LCBfQwenIIIavgprobUncalibBrierref}{0.232}
\newcommand{\LCBfQwenIIIavgprobUncalibSkillScore}{-0.257}
\newcommand{\LCBfQwenIIIavgprobUncalibACC}{0.634}
\newcommand{\LCBfQwenIIIavgprobHBECE}{0.050}
\newcommand{\LCBfQwenIIIavgprobHBASCE}{0.003}
\newcommand{\LCBfQwenIIIavgprobHBMSE}{0.168}
\newcommand{\LCBfQwenIIIavgprobHBBrierref}{0.232}
\newcommand{\LCBfQwenIIIavgprobHBSkillScore}{0.278}
\newcommand{\LCBfQwenIIIavgprobHBACC}{0.769}
\newcommand{\LCBfQwenIIIavgprobLRECE}{0.079}
\newcommand{\LCBfQwenIIIavgprobLRASCE}{0.009}
\newcommand{\LCBfQwenIIIavgprobLRMSE}{0.133}
\newcommand{\LCBfQwenIIIavgprobLRBrierref}{0.232}
\newcommand{\LCBfQwenIIIavgprobLRSkillScore}{0.425}
\newcommand{\LCBfQwenIIIavgprobLRACC}{0.824}
\newcommand{\LCBfQwenIIIavgprobLOGRECE}{0.177}
\newcommand{\LCBfQwenIIIavgprobLOGRASCE}{0.032}
\newcommand{\LCBfQwenIIIavgprobLOGRMSE}{0.177}
\newcommand{\LCBfQwenIIIavgprobLOGRBrierref}{0.232}
\newcommand{\LCBfQwenIIIavgprobLOGRSkillScore}{0.237}
\newcommand{\LCBfQwenIIIavgprobLOGRACC}{0.823}
\newcommand{\LCBfQwenIIIavgprobIGHBECE}{0.162}
\newcommand{\LCBfQwenIIIavgprobIGHBASCE}{0.053}
\newcommand{\LCBfQwenIIIavgprobIGHBMSE}{0.211}
\newcommand{\LCBfQwenIIIavgprobIGHBBrierref}{0.232}
\newcommand{\LCBfQwenIIIavgprobIGHBSkillScore}{0.090}
\newcommand{\LCBfQwenIIIavgprobIGHBACC}{0.704}
\newcommand{\LCBfQwenIIIavgprobIGLBECE}{0.069}
\newcommand{\LCBfQwenIIIavgprobIGLBASCE}{0.009}
\newcommand{\LCBfQwenIIIavgprobIGLBMSE}{0.136}
\newcommand{\LCBfQwenIIIavgprobIGLBBrierref}{0.232}
\newcommand{\LCBfQwenIIIavgprobIGLBSkillScore}{0.416}
\newcommand{\LCBfQwenIIIavgprobIGLBACC}{0.789}

\newcommand{\LCBQwenIIItailprobUncalibECE}{0.466}
\newcommand{\LCBQwenIIItailprobUncalibASCE}{0.233}
\newcommand{\LCBQwenIIItailprobUncalibMSE}{0.424}
\newcommand{\LCBQwenIIItailprobUncalibBrierref}{0.248}
\newcommand{\LCBQwenIIItailprobUncalibSkillScore}{-0.708}
\newcommand{\LCBQwenIIItailprobUncalibACC}{0.477}
\newcommand{\LCBQwenIIItailprobHBECE}{0.019}
\newcommand{\LCBQwenIIItailprobHBASCE}{0.000}
\newcommand{\LCBQwenIIItailprobHBMSE}{0.193}
\newcommand{\LCBQwenIIItailprobHBBrierref}{0.248}
\newcommand{\LCBQwenIIItailprobHBSkillScore}{0.221}
\newcommand{\LCBQwenIIItailprobHBACC}{0.688}
\newcommand{\LCBQwenIIItailprobLRECE}{0.075}
\newcommand{\LCBQwenIIItailprobLRASCE}{0.010}
\newcommand{\LCBQwenIIItailprobLRMSE}{0.136}
\newcommand{\LCBQwenIIItailprobLRBrierref}{0.248}
\newcommand{\LCBQwenIIItailprobLRSkillScore}{0.454}
\newcommand{\LCBQwenIIItailprobLRACC}{0.811}
\newcommand{\LCBQwenIIItailprobLOGRECE}{0.177}
\newcommand{\LCBQwenIIItailprobLOGRASCE}{0.032}
\newcommand{\LCBQwenIIItailprobLOGRMSE}{0.177}
\newcommand{\LCBQwenIIItailprobLOGRBrierref}{0.248}
\newcommand{\LCBQwenIIItailprobLOGRSkillScore}{0.289}
\newcommand{\LCBQwenIIItailprobLOGRACC}{0.823}
\newcommand{\LCBQwenIIItailprobIGHBECE}{0.134}
\newcommand{\LCBQwenIIItailprobIGHBASCE}{0.042}
\newcommand{\LCBQwenIIItailprobIGHBMSE}{0.200}
\newcommand{\LCBQwenIIItailprobIGHBBrierref}{0.248}
\newcommand{\LCBQwenIIItailprobIGHBSkillScore}{0.195}
\newcommand{\LCBQwenIIItailprobIGHBACC}{0.713}
\newcommand{\LCBQwenIIItailprobIGLBECE}{0.172}
\newcommand{\LCBQwenIIItailprobIGLBASCE}{0.033}
\newcommand{\LCBQwenIIItailprobIGLBMSE}{0.183}
\newcommand{\LCBQwenIIItailprobIGLBBrierref}{0.248}
\newcommand{\LCBQwenIIItailprobIGLBSkillScore}{0.264}
\newcommand{\LCBQwenIIItailprobIGLBACC}{0.806}

\newcommand{\LCBfQwenIIItailprobUncalibECE}{0.346}
\newcommand{\LCBfQwenIIItailprobUncalibASCE}{0.124}
\newcommand{\LCBfQwenIIItailprobUncalibMSE}{0.342}
\newcommand{\LCBfQwenIIItailprobUncalibBrierref}{0.232}
\newcommand{\LCBfQwenIIItailprobUncalibSkillScore}{-0.474}
\newcommand{\LCBfQwenIIItailprobUncalibACC}{0.635}
\newcommand{\LCBfQwenIIItailprobHBECE}{0.025}
\newcommand{\LCBfQwenIIItailprobHBASCE}{0.001}
\newcommand{\LCBfQwenIIItailprobHBMSE}{0.221}
\newcommand{\LCBfQwenIIItailprobHBBrierref}{0.232}
\newcommand{\LCBfQwenIIItailprobHBSkillScore}{0.046}
\newcommand{\LCBfQwenIIItailprobHBACC}{0.665}
\newcommand{\LCBfQwenIIItailprobLRECE}{0.080}
\newcommand{\LCBfQwenIIItailprobLRASCE}{0.009}
\newcommand{\LCBfQwenIIItailprobLRMSE}{0.141}
\newcommand{\LCBfQwenIIItailprobLRBrierref}{0.232}
\newcommand{\LCBfQwenIIItailprobLRSkillScore}{0.394}
\newcommand{\LCBfQwenIIItailprobLRACC}{0.811}
\newcommand{\LCBfQwenIIItailprobLOGRECE}{0.204}
\newcommand{\LCBfQwenIIItailprobLOGRASCE}{0.045}
\newcommand{\LCBfQwenIIItailprobLOGRMSE}{0.204}
\newcommand{\LCBfQwenIIItailprobLOGRBrierref}{0.232}
\newcommand{\LCBfQwenIIItailprobLOGRSkillScore}{0.119}
\newcommand{\LCBfQwenIIItailprobLOGRACC}{0.796}
\newcommand{\LCBfQwenIIItailprobIGHBECE}{0.105}
\newcommand{\LCBfQwenIIItailprobIGHBASCE}{0.030}
\newcommand{\LCBfQwenIIItailprobIGHBMSE}{0.209}
\newcommand{\LCBfQwenIIItailprobIGHBBrierref}{0.232}
\newcommand{\LCBfQwenIIItailprobIGHBSkillScore}{0.100}
\newcommand{\LCBfQwenIIItailprobIGHBACC}{0.690}
\newcommand{\LCBfQwenIIItailprobIGLBECE}{0.093}
\newcommand{\LCBfQwenIIItailprobIGLBASCE}{0.011}
\newcommand{\LCBfQwenIIItailprobIGLBMSE}{0.149}
\newcommand{\LCBfQwenIIItailprobIGLBBrierref}{0.232}
\newcommand{\LCBfQwenIIItailprobIGLBSkillScore}{0.358}
\newcommand{\LCBfQwenIIItailprobIGLBACC}{0.776}



\newcommand{\LCBGPTOSSavgprobUncalibECE}{0.199}
\newcommand{\LCBGPTOSSavgprobUncalibASCE}{0.079}
\newcommand{\LCBGPTOSSavgprobUncalibMSE}{0.269}
\newcommand{\LCBGPTOSSavgprobUncalibBrierref}{0.250}
\newcommand{\LCBGPTOSSavgprobUncalibSkillScore}{-0.075}
\newcommand{\LCBGPTOSSavgprobUncalibACC}{0.511}
\newcommand{\LCBGPTOSSavgprobPLATTECE}{0.114}
\newcommand{\LCBGPTOSSavgprobPLATTASCE}{0.021}
\newcommand{\LCBGPTOSSavgprobPLATTMSE}{0.203}
\newcommand{\LCBGPTOSSavgprobPLATTBrierref}{0.250}
\newcommand{\LCBGPTOSSavgprobPLATTSkillScore}{0.187}
\newcommand{\LCBGPTOSSavgprobPLATTACC}{0.707}
\newcommand{\LCBGPTOSSavgprobHBECE}{0.040}
\newcommand{\LCBGPTOSSavgprobHBASCE}{0.003}
\newcommand{\LCBGPTOSSavgprobHBMSE}{0.191}
\newcommand{\LCBGPTOSSavgprobHBBrierref}{0.250}
\newcommand{\LCBGPTOSSavgprobHBSkillScore}{0.237}
\newcommand{\LCBGPTOSSavgprobHBACC}{0.717}
\newcommand{\LCBGPTOSSavgprobLRECE}{0.080}
\newcommand{\LCBGPTOSSavgprobLRASCE}{0.015}
\newcommand{\LCBGPTOSSavgprobLRMSE}{0.070}
\newcommand{\LCBGPTOSSavgprobLRBrierref}{0.250}
\newcommand{\LCBGPTOSSavgprobLRSkillScore}{0.721}
\newcommand{\LCBGPTOSSavgprobLRACC}{0.924}
\newcommand{\LCBGPTOSSavgprobLOGRECE}{0.067}
\newcommand{\LCBGPTOSSavgprobLOGRASCE}{0.007}
\newcommand{\LCBGPTOSSavgprobLOGRMSE}{0.067}
\newcommand{\LCBGPTOSSavgprobLOGRBrierref}{0.250}
\newcommand{\LCBGPTOSSavgprobLOGRSkillScore}{0.733}
\newcommand{\LCBGPTOSSavgprobLOGRACC}{0.933}
\newcommand{\LCBGPTOSSavgprobIGHBECE}{0.092}
\newcommand{\LCBGPTOSSavgprobIGHBASCE}{0.021}
\newcommand{\LCBGPTOSSavgprobIGHBMSE}{0.147}
\newcommand{\LCBGPTOSSavgprobIGHBBrierref}{0.250}
\newcommand{\LCBGPTOSSavgprobIGHBSkillScore}{0.412}
\newcommand{\LCBGPTOSSavgprobIGHBACC}{0.814}
\newcommand{\LCBGPTOSSavgprobIGLBECE}{0.027}
\newcommand{\LCBGPTOSSavgprobIGLBASCE}{0.002}
\newcommand{\LCBGPTOSSavgprobIGLBMSE}{0.059}
\newcommand{\LCBGPTOSSavgprobIGLBBrierref}{0.250}
\newcommand{\LCBGPTOSSavgprobIGLBSkillScore}{0.764}
\newcommand{\LCBGPTOSSavgprobIGLBACC}{0.930}

\newcommand{\LCBfGPTOSSavgprobUncalibECE}{0.176}
\newcommand{\LCBfGPTOSSavgprobUncalibASCE}{0.032}
\newcommand{\LCBfGPTOSSavgprobUncalibMSE}{0.114}
\newcommand{\LCBfGPTOSSavgprobUncalibBrierref}{0.083}
\newcommand{\LCBfGPTOSSavgprobUncalibSkillScore}{-0.363}
\newcommand{\LCBfGPTOSSavgprobUncalibACC}{0.908}
\newcommand{\LCBfGPTOSSavgprobPLATTECE}{0.008}
\newcommand{\LCBfGPTOSSavgprobPLATTASCE}{0.000}
\newcommand{\LCBfGPTOSSavgprobPLATTMSE}{0.083}
\newcommand{\LCBfGPTOSSavgprobPLATTBrierref}{0.083}
\newcommand{\LCBfGPTOSSavgprobPLATTSkillScore}{0.004}
\newcommand{\LCBfGPTOSSavgprobPLATTACC}{0.908}
\newcommand{\LCBfGPTOSSavgprobHBECE}{0.044}
\newcommand{\LCBfGPTOSSavgprobHBASCE}{0.002}
\newcommand{\LCBfGPTOSSavgprobHBMSE}{0.084}
\newcommand{\LCBfGPTOSSavgprobHBBrierref}{0.083}
\newcommand{\LCBfGPTOSSavgprobHBSkillScore}{-0.010}
\newcommand{\LCBfGPTOSSavgprobHBACC}{0.908}
\newcommand{\LCBfGPTOSSavgprobLRECE}{0.037}
\newcommand{\LCBfGPTOSSavgprobLRASCE}{0.003}
\newcommand{\LCBfGPTOSSavgprobLRMSE}{0.079}
\newcommand{\LCBfGPTOSSavgprobLRBrierref}{0.083}
\newcommand{\LCBfGPTOSSavgprobLRSkillScore}{0.048}
\newcommand{\LCBfGPTOSSavgprobLRACC}{0.908}
\newcommand{\LCBfGPTOSSavgprobLOGRECE}{0.092}
\newcommand{\LCBfGPTOSSavgprobLOGRASCE}{0.008}
\newcommand{\LCBfGPTOSSavgprobLOGRMSE}{0.092}
\newcommand{\LCBfGPTOSSavgprobLOGRBrierref}{0.083}
\newcommand{\LCBfGPTOSSavgprobLOGRSkillScore}{-0.101}
\newcommand{\LCBfGPTOSSavgprobLOGRACC}{0.908}
\newcommand{\LCBfGPTOSSavgprobIGHBECE}{0.176}
\newcommand{\LCBfGPTOSSavgprobIGHBASCE}{0.032}
\newcommand{\LCBfGPTOSSavgprobIGHBMSE}{0.114}
\newcommand{\LCBfGPTOSSavgprobIGHBBrierref}{0.083}
\newcommand{\LCBfGPTOSSavgprobIGHBSkillScore}{-0.363}
\newcommand{\LCBfGPTOSSavgprobIGHBACC}{0.908}
\newcommand{\LCBfGPTOSSavgprobIGLBECE}{0.058}
\newcommand{\LCBfGPTOSSavgprobIGLBASCE}{0.006}
\newcommand{\LCBfGPTOSSavgprobIGLBMSE}{0.080}
\newcommand{\LCBfGPTOSSavgprobIGLBBrierref}{0.083}
\newcommand{\LCBfGPTOSSavgprobIGLBSkillScore}{0.047}
\newcommand{\LCBfGPTOSSavgprobIGLBACC}{0.908}

\newcommand{\LCBfGPTOSStailprobUncalibECE}{0.058}
\newcommand{\LCBfGPTOSStailprobUncalibASCE}{0.004}
\newcommand{\LCBfGPTOSStailprobUncalibMSE}{0.087}
\newcommand{\LCBfGPTOSStailprobUncalibBrierref}{0.083}
\newcommand{\LCBfGPTOSStailprobUncalibSkillScore}{-0.042}
\newcommand{\LCBfGPTOSStailprobUncalibACC}{0.908}
\newcommand{\LCBfGPTOSStailprobPLATTECE}{0.008}
\newcommand{\LCBfGPTOSStailprobPLATTASCE}{0.000}
\newcommand{\LCBfGPTOSStailprobPLATTMSE}{0.083}
\newcommand{\LCBfGPTOSStailprobPLATTBrierref}{0.083}
\newcommand{\LCBfGPTOSStailprobPLATTSkillScore}{-0.000}
\newcommand{\LCBfGPTOSStailprobPLATTACC}{0.908}
\newcommand{\LCBfGPTOSStailprobHBECE}{0.008}
\newcommand{\LCBfGPTOSStailprobHBASCE}{0.000}
\newcommand{\LCBfGPTOSStailprobHBMSE}{0.084}
\newcommand{\LCBfGPTOSStailprobHBBrierref}{0.083}
\newcommand{\LCBfGPTOSStailprobHBSkillScore}{-0.003}
\newcommand{\LCBfGPTOSStailprobHBACC}{0.907}
\newcommand{\LCBfGPTOSStailprobLRECE}{0.037}
\newcommand{\LCBfGPTOSStailprobLRASCE}{0.002}
\newcommand{\LCBfGPTOSStailprobLRMSE}{0.080}
\newcommand{\LCBfGPTOSStailprobLRBrierref}{0.083}
\newcommand{\LCBfGPTOSStailprobLRSkillScore}{0.037}
\newcommand{\LCBfGPTOSStailprobLRACC}{0.908}
\newcommand{\LCBfGPTOSStailprobLOGRECE}{0.092}
\newcommand{\LCBfGPTOSStailprobLOGRASCE}{0.008}
\newcommand{\LCBfGPTOSStailprobLOGRMSE}{0.092}
\newcommand{\LCBfGPTOSStailprobLOGRBrierref}{0.083}
\newcommand{\LCBfGPTOSStailprobLOGRSkillScore}{-0.101}
\newcommand{\LCBfGPTOSStailprobLOGRACC}{0.908}
\newcommand{\LCBfGPTOSStailprobIGHBECE}{0.058}
\newcommand{\LCBfGPTOSStailprobIGHBASCE}{0.004}
\newcommand{\LCBfGPTOSStailprobIGHBMSE}{0.087}
\newcommand{\LCBfGPTOSStailprobIGHBBrierref}{0.083}
\newcommand{\LCBfGPTOSStailprobIGHBSkillScore}{-0.042}
\newcommand{\LCBfGPTOSStailprobIGHBACC}{0.908}
\newcommand{\LCBfGPTOSStailprobIGLBECE}{0.057}
\newcommand{\LCBfGPTOSStailprobIGLBASCE}{0.005}
\newcommand{\LCBfGPTOSStailprobIGLBMSE}{0.086}
\newcommand{\LCBfGPTOSStailprobIGLBBrierref}{0.083}
\newcommand{\LCBfGPTOSStailprobIGLBSkillScore}{-0.025}
\newcommand{\LCBfGPTOSStailprobIGLBACC}{0.908}

\newcommand{\LCBfGPTOSScodeprobUncalibECE}{0.038}
\newcommand{\LCBfGPTOSScodeprobUncalibASCE}{0.003}
\newcommand{\LCBfGPTOSScodeprobUncalibMSE}{0.085}
\newcommand{\LCBfGPTOSScodeprobUncalibBrierref}{0.083}
\newcommand{\LCBfGPTOSScodeprobUncalibSkillScore}{-0.020}
\newcommand{\LCBfGPTOSScodeprobUncalibACC}{0.908}
\newcommand{\LCBfGPTOSScodeprobPLATTECE}{0.008}
\newcommand{\LCBfGPTOSScodeprobPLATTASCE}{0.000}
\newcommand{\LCBfGPTOSScodeprobPLATTMSE}{0.083}
\newcommand{\LCBfGPTOSScodeprobPLATTBrierref}{0.083}
\newcommand{\LCBfGPTOSScodeprobPLATTSkillScore}{-0.000}
\newcommand{\LCBfGPTOSScodeprobPLATTACC}{0.908}
\newcommand{\LCBfGPTOSScodeprobHBECE}{0.004}
\newcommand{\LCBfGPTOSScodeprobHBASCE}{0.001}
\newcommand{\LCBfGPTOSScodeprobHBMSE}{0.084}
\newcommand{\LCBfGPTOSScodeprobHBBrierref}{0.083}
\newcommand{\LCBfGPTOSScodeprobHBSkillScore}{-0.002}
\newcommand{\LCBfGPTOSScodeprobHBACC}{0.908}
\newcommand{\LCBfGPTOSScodeprobLRECE}{0.038}
\newcommand{\LCBfGPTOSScodeprobLRASCE}{0.002}
\newcommand{\LCBfGPTOSScodeprobLRMSE}{0.080}
\newcommand{\LCBfGPTOSScodeprobLRBrierref}{0.083}
\newcommand{\LCBfGPTOSScodeprobLRSkillScore}{0.037}
\newcommand{\LCBfGPTOSScodeprobLRACC}{0.908}
\newcommand{\LCBfGPTOSScodeprobLOGRECE}{0.092}
\newcommand{\LCBfGPTOSScodeprobLOGRASCE}{0.008}
\newcommand{\LCBfGPTOSScodeprobLOGRMSE}{0.092}
\newcommand{\LCBfGPTOSScodeprobLOGRBrierref}{0.083}
\newcommand{\LCBfGPTOSScodeprobLOGRSkillScore}{-0.101}
\newcommand{\LCBfGPTOSScodeprobLOGRACC}{0.908}
\newcommand{\LCBfGPTOSScodeprobIGHBECE}{0.038}
\newcommand{\LCBfGPTOSScodeprobIGHBASCE}{0.003}
\newcommand{\LCBfGPTOSScodeprobIGHBMSE}{0.085}
\newcommand{\LCBfGPTOSScodeprobIGHBBrierref}{0.083}
\newcommand{\LCBfGPTOSScodeprobIGHBSkillScore}{-0.020}
\newcommand{\LCBfGPTOSScodeprobIGHBACC}{0.908}
\newcommand{\LCBfGPTOSScodeprobIGLBECE}{0.037}
\newcommand{\LCBfGPTOSScodeprobIGLBASCE}{0.002}
\newcommand{\LCBfGPTOSScodeprobIGLBMSE}{0.085}
\newcommand{\LCBfGPTOSScodeprobIGLBBrierref}{0.083}
\newcommand{\LCBfGPTOSScodeprobIGLBSkillScore}{-0.015}
\newcommand{\LCBfGPTOSScodeprobIGLBACC}{0.908}
 \newcommand{\MEQwenIIIavgprobUncalibgASCEpromptlenlow}{ 0.085}
\newcommand{\MEQwenIIIavgprobUncalibgASCEloclow}{ 0.051}
 \newcommand{\MEQwenIIIavgprobUncalibgASCElenlow}{ 0.041}
\newcommand{\MEQwenIIIavgprobPLATTgASCEpromptlenlow}{ 0.003}
 \newcommand{\MEQwenIIIavgprobPLATTgASCEloclow}{ 0.007}
 \newcommand{\MEQwenIIIavgprobPLATTgASCElenlow}{ 0.003}
\newcommand{\MEQwenIIIavgprobHBgASCEpromptlenlow}{ 0.002}
 \newcommand{\MEQwenIIIavgprobHBgASCEloclow}{ 0.006}
 \newcommand{\MEQwenIIIavgprobHBgASCElenlow}{ 0.004}
 \newcommand{\MEQwenIIIavgprobLRgASCEpromptlenlow}{ 0.003}
 \newcommand{\MEQwenIIIavgprobLRgASCEloclow}{ 0.003}
 \newcommand{\MEQwenIIIavgprobLRgASCElenlow}{ 0.003}
 \newcommand{\MEQwenIIIavgprobLOGRgASCEpromptlenlow}{ 0.057}
\newcommand{\MEQwenIIIavgprobLOGRgASCEloclow}{ 0.055}
 \newcommand{\MEQwenIIIavgprobLOGRgASCElenlow}{ 0.038}
 \newcommand{\MEQwenIIIavgprobIGHBgASCEpromptlenlow}{ 0.046}
\newcommand{\MEQwenIIIavgprobIGHBgASCEloclow}{ 0.042}
 \newcommand{\MEQwenIIIavgprobIGHBgASCElenlow}{ 0.031}
 \newcommand{\MEQwenIIIavgprobIGLBgASCEpromptlenlow}{ 0.002}
\newcommand{\MEQwenIIIavgprobIGLBgASCEloclow}{ 0.006}
 \newcommand{\MEQwenIIIavgprobIGLBgASCElenlow}{ 0.003}

\newcommand{\MEQwenIIIavgprobUncalibECE}{ 0.295}
\newcommand{\MEQwenIIIavgprobUncalibASCE}{ 0.096}
\newcommand{\MEQwenIIIavgprobUncalibMSE}{ 0.296}
\newcommand{\MEQwenIIIavgprobUncalibBrierref}{ 0.238}
\newcommand{\MEQwenIIIavgprobUncalibSkillScore}{ -0.243}
\newcommand{\MEQwenIIIavgprobUncalibACC}{ 0.613}
\newcommand{\MEQwenIIIavgprobUncalibgASCEcompeasy}{ 0.052}
\newcommand{\MEQwenIIIavgprobUncalibgASCEcompmedium}{ 0.119}
\newcommand{\MEQwenIIIavgprobUncalibgASCEcomphard}{ 0.261}
\newcommand{\MEQwenIIIavgprobUncalibgASCEpromptlenhigh}{ 0.109}
\newcommand{\MEQwenIIIavgprobUncalibgASCElochigh}{ 0.149}
\newcommand{\MEQwenIIIavgprobUncalibgASCElenhigh}{ 0.156}
\newcommand{\MEQwenIIIavgprobUncalibgASCElangC}{ 0.130}
\newcommand{\MEQwenIIIavgprobUncalibgASCElangCPP}{ 0.196}
\newcommand{\MEQwenIIIavgprobUncalibgASCElangCoffeeScript}{ 0.017}
\newcommand{\MEQwenIIIavgprobUncalibgASCElangCommonLisp}{ 0.124}
\newcommand{\MEQwenIIIavgprobUncalibgASCElangDart}{ 0.292}
\newcommand{\MEQwenIIIavgprobUncalibgASCElangElixir}{ 0.117}
\newcommand{\MEQwenIIIavgprobUncalibgASCElangEmacsLisp}{ 0.350}
\newcommand{\MEQwenIIIavgprobUncalibgASCElangFortran}{ 0.091}
\newcommand{\MEQwenIIIavgprobUncalibgASCElangGo}{ 0.099}
\newcommand{\MEQwenIIIavgprobUncalibgASCElangGroovy}{ 0.033}
\newcommand{\MEQwenIIIavgprobUncalibgASCElangHTML}{ 0.440}
\newcommand{\MEQwenIIIavgprobUncalibgASCElangHaskell}{ 0.087}
\newcommand{\MEQwenIIIavgprobUncalibgASCElangJSON}{ 0.067}
\newcommand{\MEQwenIIIavgprobUncalibgASCElangJava}{ 0.035}
\newcommand{\MEQwenIIIavgprobUncalibgASCElangJavaScript}{ 0.164}
\newcommand{\MEQwenIIIavgprobUncalibgASCElangJulia}{ 0.061}
\newcommand{\MEQwenIIIavgprobUncalibgASCElangKotlin}{ 0.020}
\newcommand{\MEQwenIIIavgprobUncalibgASCElangLua}{ 0.119}
\newcommand{\MEQwenIIIavgprobUncalibgASCElangPHP}{ 0.092}
\newcommand{\MEQwenIIIavgprobUncalibgASCElangPascal}{ 0.145}
\newcommand{\MEQwenIIIavgprobUncalibgASCElangPerl}{ 0.225}
\newcommand{\MEQwenIIIavgprobUncalibgASCElangPowerShell}{ 0.032}
\newcommand{\MEQwenIIIavgprobUncalibgASCElangPython}{ 0.064}
\newcommand{\MEQwenIIIavgprobUncalibgASCElangR}{ 0.212}
\newcommand{\MEQwenIIIavgprobUncalibgASCElangRacket}{ 0.165}
\newcommand{\MEQwenIIIavgprobUncalibgASCElangRuby}{ 0.034}
\newcommand{\MEQwenIIIavgprobUncalibgASCElangRust}{ 0.556}
\newcommand{\MEQwenIIIavgprobUncalibgASCElangScala}{ 0.230}
\newcommand{\MEQwenIIIavgprobUncalibgASCElangScheme}{ 0.101}
\newcommand{\MEQwenIIIavgprobUncalibgASCElangShell}{ 0.077}
\newcommand{\MEQwenIIIavgprobUncalibgASCElangSwift}{ 0.027}
\newcommand{\MEQwenIIIavgprobUncalibgASCElangTcl}{ 0.044}
\newcommand{\MEQwenIIIavgprobUncalibgASCElangTypeScript}{ 0.147}
\newcommand{\MEQwenIIIavgprobUncalibgASCElangVimScript}{ 0.169}
\newcommand{\MEQwenIIIavgprobPLATTECE}{ 0.034}
\newcommand{\MEQwenIIIavgprobPLATTASCE}{ 0.002}
\newcommand{\MEQwenIIIavgprobPLATTMSE}{ 0.200}
\newcommand{\MEQwenIIIavgprobPLATTBrierref}{ 0.238}
\newcommand{\MEQwenIIIavgprobPLATTSkillScore}{ 0.160}
\newcommand{\MEQwenIIIavgprobPLATTACC}{ 0.699}
\newcommand{\MEQwenIIIavgprobPLATTgASCEcompeasy}{ 0.005}
\newcommand{\MEQwenIIIavgprobPLATTgASCEcompmedium}{ 0.006}
\newcommand{\MEQwenIIIavgprobPLATTgASCEcomphard}{ 0.047}
\newcommand{\MEQwenIIIavgprobPLATTgASCEpromptlenhigh}{ 0.003}
\newcommand{\MEQwenIIIavgprobPLATTgASCElochigh}{ 0.006}
\newcommand{\MEQwenIIIavgprobPLATTgASCElenhigh}{ 0.006}
\newcommand{\MEQwenIIIavgprobPLATTgASCElangC}{ 0.039}
\newcommand{\MEQwenIIIavgprobPLATTgASCElangCPP}{ 0.030}
\newcommand{\MEQwenIIIavgprobPLATTgASCElangCoffeeScript}{ 0.057}
\newcommand{\MEQwenIIIavgprobPLATTgASCElangCommonLisp}{ 0.023}
\newcommand{\MEQwenIIIavgprobPLATTgASCElangDart}{ 0.061}
\newcommand{\MEQwenIIIavgprobPLATTgASCElangElixir}{ 0.017}
\newcommand{\MEQwenIIIavgprobPLATTgASCElangEmacsLisp}{ 0.116}
\newcommand{\MEQwenIIIavgprobPLATTgASCElangFortran}{ 0.012}
\newcommand{\MEQwenIIIavgprobPLATTgASCElangGo}{ 0.018}
\newcommand{\MEQwenIIIavgprobPLATTgASCElangGroovy}{ 0.022}
\newcommand{\MEQwenIIIavgprobPLATTgASCElangHTML}{ 0.192}
\newcommand{\MEQwenIIIavgprobPLATTgASCElangHaskell}{ 0.049}
\newcommand{\MEQwenIIIavgprobPLATTgASCElangJSON}{ 0.050}
\newcommand{\MEQwenIIIavgprobPLATTgASCElangJava}{ 0.058}
\newcommand{\MEQwenIIIavgprobPLATTgASCElangJavaScript}{ 0.033}
\newcommand{\MEQwenIIIavgprobPLATTgASCElangJulia}{ 0.029}
\newcommand{\MEQwenIIIavgprobPLATTgASCElangKotlin}{ 0.071}
\newcommand{\MEQwenIIIavgprobPLATTgASCElangLua}{ 0.031}
\newcommand{\MEQwenIIIavgprobPLATTgASCElangPHP}{ 0.035}
\newcommand{\MEQwenIIIavgprobPLATTgASCElangPascal}{ 0.011}
\newcommand{\MEQwenIIIavgprobPLATTgASCElangPerl}{ 0.044}
\newcommand{\MEQwenIIIavgprobPLATTgASCElangPowerShell}{ 0.036}
\newcommand{\MEQwenIIIavgprobPLATTgASCElangPython}{ 0.051}
\newcommand{\MEQwenIIIavgprobPLATTgASCElangR}{ 0.064}
\newcommand{\MEQwenIIIavgprobPLATTgASCElangRacket}{ 0.036}
\newcommand{\MEQwenIIIavgprobPLATTgASCElangRuby}{ 0.085}
\newcommand{\MEQwenIIIavgprobPLATTgASCElangRust}{ 0.250}
\newcommand{\MEQwenIIIavgprobPLATTgASCElangScala}{ 0.079}
\newcommand{\MEQwenIIIavgprobPLATTgASCElangScheme}{ 0.025}
\newcommand{\MEQwenIIIavgprobPLATTgASCElangShell}{ 0.026}
\newcommand{\MEQwenIIIavgprobPLATTgASCElangSwift}{ 0.039}
\newcommand{\MEQwenIIIavgprobPLATTgASCElangTcl}{ 0.022}
\newcommand{\MEQwenIIIavgprobPLATTgASCElangTypeScript}{ 0.038}
\newcommand{\MEQwenIIIavgprobPLATTgASCElangVimScript}{ 0.080}
\newcommand{\MEQwenIIIavgprobHBECE}{ 0.022}
\newcommand{\MEQwenIIIavgprobHBASCE}{ 0.001}
\newcommand{\MEQwenIIIavgprobHBMSE}{ 0.200}
\newcommand{\MEQwenIIIavgprobHBBrierref}{ 0.238}
\newcommand{\MEQwenIIIavgprobHBSkillScore}{ 0.160}
\newcommand{\MEQwenIIIavgprobHBACC}{ 0.704}
\newcommand{\MEQwenIIIavgprobHBgASCEcompeasy}{ 0.006}
\newcommand{\MEQwenIIIavgprobHBgASCEcompmedium}{ 0.001}
\newcommand{\MEQwenIIIavgprobHBgASCEcomphard}{ 0.034}
\newcommand{\MEQwenIIIavgprobHBgASCEpromptlenhigh}{ 0.001}
\newcommand{\MEQwenIIIavgprobHBgASCElochigh}{ 0.003}
\newcommand{\MEQwenIIIavgprobHBgASCElenhigh}{ 0.003}
\newcommand{\MEQwenIIIavgprobHBgASCElangC}{ 0.027}
\newcommand{\MEQwenIIIavgprobHBgASCElangCPP}{ 0.025}
\newcommand{\MEQwenIIIavgprobHBgASCElangCoffeeScript}{ 0.061}
\newcommand{\MEQwenIIIavgprobHBgASCElangCommonLisp}{ 0.008}
\newcommand{\MEQwenIIIavgprobHBgASCElangDart}{ 0.042}
\newcommand{\MEQwenIIIavgprobHBgASCElangElixir}{ 0.014}
\newcommand{\MEQwenIIIavgprobHBgASCElangEmacsLisp}{ 0.088}
\newcommand{\MEQwenIIIavgprobHBgASCElangFortran}{ 0.011}
\newcommand{\MEQwenIIIavgprobHBgASCElangGo}{ 0.016}
\newcommand{\MEQwenIIIavgprobHBgASCElangGroovy}{ 0.016}
\newcommand{\MEQwenIIIavgprobHBgASCElangHTML}{ 0.173}
\newcommand{\MEQwenIIIavgprobHBgASCElangHaskell}{ 0.019}
\newcommand{\MEQwenIIIavgprobHBgASCElangJSON}{ 0.021}
\newcommand{\MEQwenIIIavgprobHBgASCElangJava}{ 0.041}
\newcommand{\MEQwenIIIavgprobHBgASCElangJavaScript}{ 0.037}
\newcommand{\MEQwenIIIavgprobHBgASCElangJulia}{ 0.029}
\newcommand{\MEQwenIIIavgprobHBgASCElangKotlin}{ 0.066}
\newcommand{\MEQwenIIIavgprobHBgASCElangLua}{ 0.034}
\newcommand{\MEQwenIIIavgprobHBgASCElangPHP}{ 0.020}
\newcommand{\MEQwenIIIavgprobHBgASCElangPascal}{ 0.015}
\newcommand{\MEQwenIIIavgprobHBgASCElangPerl}{ 0.035}
\newcommand{\MEQwenIIIavgprobHBgASCElangPowerShell}{ 0.033}
\newcommand{\MEQwenIIIavgprobHBgASCElangPython}{ 0.029}
\newcommand{\MEQwenIIIavgprobHBgASCElangR}{ 0.041}
\newcommand{\MEQwenIIIavgprobHBgASCElangRacket}{ 0.024}
\newcommand{\MEQwenIIIavgprobHBgASCElangRuby}{ 0.073}
\newcommand{\MEQwenIIIavgprobHBgASCElangRust}{ 0.223}
\newcommand{\MEQwenIIIavgprobHBgASCElangScala}{ 0.054}
\newcommand{\MEQwenIIIavgprobHBgASCElangScheme}{ 0.014}
\newcommand{\MEQwenIIIavgprobHBgASCElangShell}{ 0.013}
\newcommand{\MEQwenIIIavgprobHBgASCElangSwift}{ 0.031}
\newcommand{\MEQwenIIIavgprobHBgASCElangTcl}{ 0.017}
\newcommand{\MEQwenIIIavgprobHBgASCElangTypeScript}{ 0.013}
\newcommand{\MEQwenIIIavgprobHBgASCElangVimScript}{ 0.058}
\newcommand{\MEQwenIIIavgprobLRECE}{ 0.047}
\newcommand{\MEQwenIIIavgprobLRASCE}{ 0.003}
\newcommand{\MEQwenIIIavgprobLRMSE}{ 0.167}
\newcommand{\MEQwenIIIavgprobLRBrierref}{ 0.238}
\newcommand{\MEQwenIIIavgprobLRSkillScore}{ 0.297}
\newcommand{\MEQwenIIIavgprobLRACC}{ 0.763}
\newcommand{\MEQwenIIIavgprobLRgASCEcompeasy}{ 0.003}
\newcommand{\MEQwenIIIavgprobLRgASCEcompmedium}{ 0.011}
\newcommand{\MEQwenIIIavgprobLRgASCEcomphard}{ 0.008}
\newcommand{\MEQwenIIIavgprobLRgASCEpromptlenhigh}{ 0.004}
\newcommand{\MEQwenIIIavgprobLRgASCElochigh}{ 0.007}
\newcommand{\MEQwenIIIavgprobLRgASCElenhigh}{ 0.006}
\newcommand{\MEQwenIIIavgprobLRgASCElangC}{ 0.026}
\newcommand{\MEQwenIIIavgprobLRgASCElangCPP}{ 0.024}
\newcommand{\MEQwenIIIavgprobLRgASCElangCoffeeScript}{ 0.032}
\newcommand{\MEQwenIIIavgprobLRgASCElangCommonLisp}{ 0.051}
\newcommand{\MEQwenIIIavgprobLRgASCElangDart}{ 0.048}
\newcommand{\MEQwenIIIavgprobLRgASCElangElixir}{ 0.022}
\newcommand{\MEQwenIIIavgprobLRgASCElangEmacsLisp}{ 0.011}
\newcommand{\MEQwenIIIavgprobLRgASCElangFortran}{ 0.018}
\newcommand{\MEQwenIIIavgprobLRgASCElangGo}{ 0.029}
\newcommand{\MEQwenIIIavgprobLRgASCElangGroovy}{ 0.018}
\newcommand{\MEQwenIIIavgprobLRgASCElangHTML}{ 0.017}
\newcommand{\MEQwenIIIavgprobLRgASCElangHaskell}{ 0.021}
\newcommand{\MEQwenIIIavgprobLRgASCElangJSON}{ 0.041}
\newcommand{\MEQwenIIIavgprobLRgASCElangJava}{ 0.021}
\newcommand{\MEQwenIIIavgprobLRgASCElangJavaScript}{ 0.029}
\newcommand{\MEQwenIIIavgprobLRgASCElangJulia}{ 0.033}
\newcommand{\MEQwenIIIavgprobLRgASCElangKotlin}{ 0.024}
\newcommand{\MEQwenIIIavgprobLRgASCElangLua}{ 0.032}
\newcommand{\MEQwenIIIavgprobLRgASCElangPHP}{ 0.031}
\newcommand{\MEQwenIIIavgprobLRgASCElangPascal}{ 0.038}
\newcommand{\MEQwenIIIavgprobLRgASCElangPerl}{ 0.028}
\newcommand{\MEQwenIIIavgprobLRgASCElangPowerShell}{ 0.025}
\newcommand{\MEQwenIIIavgprobLRgASCElangPython}{ 0.037}
\newcommand{\MEQwenIIIavgprobLRgASCElangR}{ 0.035}
\newcommand{\MEQwenIIIavgprobLRgASCElangRacket}{ 0.032}
\newcommand{\MEQwenIIIavgprobLRgASCElangRuby}{ 0.037}
\newcommand{\MEQwenIIIavgprobLRgASCElangRust}{ 0.037}
\newcommand{\MEQwenIIIavgprobLRgASCElangScala}{ 0.016}
\newcommand{\MEQwenIIIavgprobLRgASCElangScheme}{ 0.021}
\newcommand{\MEQwenIIIavgprobLRgASCElangShell}{ 0.023}
\newcommand{\MEQwenIIIavgprobLRgASCElangSwift}{ 0.017}
\newcommand{\MEQwenIIIavgprobLRgASCElangTcl}{ 0.017}
\newcommand{\MEQwenIIIavgprobLRgASCElangTypeScript}{ 0.038}
\newcommand{\MEQwenIIIavgprobLRgASCElangVimScript}{ 0.106}
\newcommand{\MEQwenIIIavgprobLOGRECE}{ 0.229}
\newcommand{\MEQwenIIIavgprobLOGRASCE}{ 0.053}
\newcommand{\MEQwenIIIavgprobLOGRMSE}{ 0.229}
\newcommand{\MEQwenIIIavgprobLOGRBrierref}{ 0.238}
\newcommand{\MEQwenIIIavgprobLOGRSkillScore}{ 0.038}
\newcommand{\MEQwenIIIavgprobLOGRACC}{ 0.771}
\newcommand{\MEQwenIIIavgprobLOGRgASCEcompeasy}{ 0.045}
\newcommand{\MEQwenIIIavgprobLOGRgASCEcompmedium}{ 0.074}
\newcommand{\MEQwenIIIavgprobLOGRgASCEcomphard}{ 0.061}
\newcommand{\MEQwenIIIavgprobLOGRgASCEpromptlenhigh}{ 0.048}
\newcommand{\MEQwenIIIavgprobLOGRgASCElochigh}{ 0.050}
\newcommand{\MEQwenIIIavgprobLOGRgASCElenhigh}{ 0.070}
\newcommand{\MEQwenIIIavgprobLOGRgASCElangC}{ 0.017}
\newcommand{\MEQwenIIIavgprobLOGRgASCElangCPP}{ 0.026}
\newcommand{\MEQwenIIIavgprobLOGRgASCElangCoffeeScript}{ 0.020}
\newcommand{\MEQwenIIIavgprobLOGRgASCElangCommonLisp}{ 0.063}
\newcommand{\MEQwenIIIavgprobLOGRgASCElangDart}{ 0.160}
\newcommand{\MEQwenIIIavgprobLOGRgASCElangElixir}{ 0.062}
\newcommand{\MEQwenIIIavgprobLOGRgASCElangEmacsLisp}{ 0.115}
\newcommand{\MEQwenIIIavgprobLOGRgASCElangFortran}{ 0.062}
\newcommand{\MEQwenIIIavgprobLOGRgASCElangGo}{ 0.023}
\newcommand{\MEQwenIIIavgprobLOGRgASCElangGroovy}{ 0.039}
\newcommand{\MEQwenIIIavgprobLOGRgASCElangHTML}{ 0.105}
\newcommand{\MEQwenIIIavgprobLOGRgASCElangHaskell}{ 0.051}
\newcommand{\MEQwenIIIavgprobLOGRgASCElangJSON}{ 0.024}
\newcommand{\MEQwenIIIavgprobLOGRgASCElangJava}{ 0.046}
\newcommand{\MEQwenIIIavgprobLOGRgASCElangJavaScript}{ 0.019}
\newcommand{\MEQwenIIIavgprobLOGRgASCElangJulia}{ 0.028}
\newcommand{\MEQwenIIIavgprobLOGRgASCElangKotlin}{ 0.014}
\newcommand{\MEQwenIIIavgprobLOGRgASCElangLua}{ 0.014}
\newcommand{\MEQwenIIIavgprobLOGRgASCElangPHP}{ 0.054}
\newcommand{\MEQwenIIIavgprobLOGRgASCElangPascal}{ 0.052}
\newcommand{\MEQwenIIIavgprobLOGRgASCElangPerl}{ 0.083}
\newcommand{\MEQwenIIIavgprobLOGRgASCElangPowerShell}{ 0.026}
\newcommand{\MEQwenIIIavgprobLOGRgASCElangPython}{ 0.076}
\newcommand{\MEQwenIIIavgprobLOGRgASCElangR}{ 0.062}
\newcommand{\MEQwenIIIavgprobLOGRgASCElangRacket}{ 0.106}
\newcommand{\MEQwenIIIavgprobLOGRgASCElangRuby}{ 0.114}
\newcommand{\MEQwenIIIavgprobLOGRgASCElangRust}{ 0.036}
\newcommand{\MEQwenIIIavgprobLOGRgASCElangScala}{ 0.140}
\newcommand{\MEQwenIIIavgprobLOGRgASCElangScheme}{ 0.064}
\newcommand{\MEQwenIIIavgprobLOGRgASCElangShell}{ 0.170}
\newcommand{\MEQwenIIIavgprobLOGRgASCElangSwift}{ 0.039}
\newcommand{\MEQwenIIIavgprobLOGRgASCElangTcl}{ 0.071}
\newcommand{\MEQwenIIIavgprobLOGRgASCElangTypeScript}{ 0.023}
\newcommand{\MEQwenIIIavgprobLOGRgASCElangVimScript}{ 0.123}
\newcommand{\MEQwenIIIavgprobIGHBECE}{ 0.191}
\newcommand{\MEQwenIIIavgprobIGHBASCE}{ 0.055}
\newcommand{\MEQwenIIIavgprobIGHBMSE}{ 0.267}
\newcommand{\MEQwenIIIavgprobIGHBBrierref}{ 0.238}
\newcommand{\MEQwenIIIavgprobIGHBSkillScore}{ -0.122}
\newcommand{\MEQwenIIIavgprobIGHBACC}{ 0.613}
\newcommand{\MEQwenIIIavgprobIGHBgASCEcompeasy}{ 0.030}
\newcommand{\MEQwenIIIavgprobIGHBgASCEcompmedium}{ 0.082}
\newcommand{\MEQwenIIIavgprobIGHBgASCEcomphard}{ 0.164}
\newcommand{\MEQwenIIIavgprobIGHBgASCEpromptlenhigh}{ 0.067}
\newcommand{\MEQwenIIIavgprobIGHBgASCElochigh}{ 0.072}
\newcommand{\MEQwenIIIavgprobIGHBgASCElenhigh}{ 0.086}
\newcommand{\MEQwenIIIavgprobIGHBgASCElangC}{ 0.086}
\newcommand{\MEQwenIIIavgprobIGHBgASCElangCPP}{ 0.137}
\newcommand{\MEQwenIIIavgprobIGHBgASCElangCoffeeScript}{ 0.031}
\newcommand{\MEQwenIIIavgprobIGHBgASCElangCommonLisp}{ 0.082}
\newcommand{\MEQwenIIIavgprobIGHBgASCElangDart}{ 0.202}
\newcommand{\MEQwenIIIavgprobIGHBgASCElangElixir}{ 0.095}
\newcommand{\MEQwenIIIavgprobIGHBgASCElangEmacsLisp}{ 0.267}
\newcommand{\MEQwenIIIavgprobIGHBgASCElangFortran}{ 0.046}
\newcommand{\MEQwenIIIavgprobIGHBgASCElangGo}{ 0.061}
\newcommand{\MEQwenIIIavgprobIGHBgASCElangGroovy}{ 0.071}
\newcommand{\MEQwenIIIavgprobIGHBgASCElangHTML}{ 0.430}
\newcommand{\MEQwenIIIavgprobIGHBgASCElangHaskell}{ 0.067}
\newcommand{\MEQwenIIIavgprobIGHBgASCElangJSON}{ 0.077}
\newcommand{\MEQwenIIIavgprobIGHBgASCElangJava}{ 0.065}
\newcommand{\MEQwenIIIavgprobIGHBgASCElangJavaScript}{ 0.100}
\newcommand{\MEQwenIIIavgprobIGHBgASCElangJulia}{ 0.081}
\newcommand{\MEQwenIIIavgprobIGHBgASCElangKotlin}{ 0.055}
\newcommand{\MEQwenIIIavgprobIGHBgASCElangLua}{ 0.070}
\newcommand{\MEQwenIIIavgprobIGHBgASCElangPHP}{ 0.076}
\newcommand{\MEQwenIIIavgprobIGHBgASCElangPascal}{ 0.103}
\newcommand{\MEQwenIIIavgprobIGHBgASCElangPerl}{ 0.185}
\newcommand{\MEQwenIIIavgprobIGHBgASCElangPowerShell}{ 0.049}
\newcommand{\MEQwenIIIavgprobIGHBgASCElangPython}{ 0.035}
\newcommand{\MEQwenIIIavgprobIGHBgASCElangR}{ 0.157}
\newcommand{\MEQwenIIIavgprobIGHBgASCElangRacket}{ 0.122}
\newcommand{\MEQwenIIIavgprobIGHBgASCElangRuby}{ 0.038}
\newcommand{\MEQwenIIIavgprobIGHBgASCElangRust}{ 0.323}
\newcommand{\MEQwenIIIavgprobIGHBgASCElangScala}{ 0.138}
\newcommand{\MEQwenIIIavgprobIGHBgASCElangScheme}{ 0.069}
\newcommand{\MEQwenIIIavgprobIGHBgASCElangShell}{ 0.034}
\newcommand{\MEQwenIIIavgprobIGHBgASCElangSwift}{ 0.042}
\newcommand{\MEQwenIIIavgprobIGHBgASCElangTcl}{ 0.020}
\newcommand{\MEQwenIIIavgprobIGHBgASCElangTypeScript}{ 0.086}
\newcommand{\MEQwenIIIavgprobIGHBgASCElangVimScript}{ 0.118}
\newcommand{\MEQwenIIIavgprobIGLBECE}{ 0.034}
\newcommand{\MEQwenIIIavgprobIGLBASCE}{ 0.002}
\newcommand{\MEQwenIIIavgprobIGLBMSE}{ 0.163}
\newcommand{\MEQwenIIIavgprobIGLBBrierref}{ 0.238}
\newcommand{\MEQwenIIIavgprobIGLBSkillScore}{ 0.317}
\newcommand{\MEQwenIIIavgprobIGLBACC}{ 0.768}
\newcommand{\MEQwenIIIavgprobIGLBgASCEcompeasy}{ 0.003}
\newcommand{\MEQwenIIIavgprobIGLBgASCEcompmedium}{ 0.005}
\newcommand{\MEQwenIIIavgprobIGLBgASCEcomphard}{ 0.007}
\newcommand{\MEQwenIIIavgprobIGLBgASCEpromptlenhigh}{ 0.002}
\newcommand{\MEQwenIIIavgprobIGLBgASCElochigh}{ 0.003}
\newcommand{\MEQwenIIIavgprobIGLBgASCElenhigh}{ 0.003}
\newcommand{\MEQwenIIIavgprobIGLBgASCElangC}{ 0.019}
\newcommand{\MEQwenIIIavgprobIGLBgASCElangCPP}{ 0.031}
\newcommand{\MEQwenIIIavgprobIGLBgASCElangCoffeeScript}{ 0.005}
\newcommand{\MEQwenIIIavgprobIGLBgASCElangCommonLisp}{ 0.074}
\newcommand{\MEQwenIIIavgprobIGLBgASCElangDart}{ 0.026}
\newcommand{\MEQwenIIIavgprobIGLBgASCElangElixir}{ 0.031}
\newcommand{\MEQwenIIIavgprobIGLBgASCElangEmacsLisp}{ 0.024}
\newcommand{\MEQwenIIIavgprobIGLBgASCElangFortran}{ 0.027}
\newcommand{\MEQwenIIIavgprobIGLBgASCElangGo}{ 0.050}
\newcommand{\MEQwenIIIavgprobIGLBgASCElangGroovy}{ 0.042}
\newcommand{\MEQwenIIIavgprobIGLBgASCElangHTML}{ 0.063}
\newcommand{\MEQwenIIIavgprobIGLBgASCElangHaskell}{ 0.033}
\newcommand{\MEQwenIIIavgprobIGLBgASCElangJSON}{ 0.035}
\newcommand{\MEQwenIIIavgprobIGLBgASCElangJava}{ 0.029}
\newcommand{\MEQwenIIIavgprobIGLBgASCElangJavaScript}{ 0.046}
\newcommand{\MEQwenIIIavgprobIGLBgASCElangJulia}{ 0.052}
\newcommand{\MEQwenIIIavgprobIGLBgASCElangKotlin}{ 0.012}
\newcommand{\MEQwenIIIavgprobIGLBgASCElangLua}{ 0.033}
\newcommand{\MEQwenIIIavgprobIGLBgASCElangPHP}{ 0.045}
\newcommand{\MEQwenIIIavgprobIGLBgASCElangPascal}{ 0.015}
\newcommand{\MEQwenIIIavgprobIGLBgASCElangPerl}{ 0.019}
\newcommand{\MEQwenIIIavgprobIGLBgASCElangPowerShell}{ 0.045}
\newcommand{\MEQwenIIIavgprobIGLBgASCElangPython}{ 0.027}
\newcommand{\MEQwenIIIavgprobIGLBgASCElangR}{ 0.029}
\newcommand{\MEQwenIIIavgprobIGLBgASCElangRacket}{ 0.065}
\newcommand{\MEQwenIIIavgprobIGLBgASCElangRuby}{ 0.014}
\newcommand{\MEQwenIIIavgprobIGLBgASCElangRust}{ 0.036}
\newcommand{\MEQwenIIIavgprobIGLBgASCElangScala}{ 0.021}
\newcommand{\MEQwenIIIavgprobIGLBgASCElangScheme}{ 0.036}
\newcommand{\MEQwenIIIavgprobIGLBgASCElangShell}{ 0.053}
\newcommand{\MEQwenIIIavgprobIGLBgASCElangSwift}{ 0.035}
\newcommand{\MEQwenIIIavgprobIGLBgASCElangTcl}{ 0.020}
\newcommand{\MEQwenIIIavgprobIGLBgASCElangTypeScript}{ 0.030}
\newcommand{\MEQwenIIIavgprobIGLBgASCElangVimScript}{ 0.049}

\newcommand{\MEDSRavgprobUncalibECE}{ 0.381}
\newcommand{\MEDSRavgprobUncalibASCE}{ 0.153}
\newcommand{\MEDSRavgprobUncalibMSE}{ 0.326}
\newcommand{\MEDSRavgprobUncalibBrierref}{ 0.195}
\newcommand{\MEDSRavgprobUncalibSkillScore}{ -0.670}
\newcommand{\MEDSRavgprobUncalibACC}{ 0.288}
\newcommand{\MEDSRavgprobPLATTECE}{ 0.048}
\newcommand{\MEDSRavgprobPLATTASCE}{ 0.005}
\newcommand{\MEDSRavgprobPLATTMSE}{ 0.177}
\newcommand{\MEDSRavgprobPLATTBrierref}{ 0.195}
\newcommand{\MEDSRavgprobPLATTSkillScore}{ 0.093}
\newcommand{\MEDSRavgprobPLATTACC}{ 0.734}
\newcommand{\MEDSRavgprobHBECE}{ 0.015}
\newcommand{\MEDSRavgprobHBASCE}{ 0.000}
\newcommand{\MEDSRavgprobHBMSE}{ 0.173}
\newcommand{\MEDSRavgprobHBBrierref}{ 0.195}
\newcommand{\MEDSRavgprobHBSkillScore}{ 0.112}
\newcommand{\MEDSRavgprobHBACC}{ 0.755}
\newcommand{\MEDSRavgprobLRECE}{ 0.090}
\newcommand{\MEDSRavgprobLRASCE}{ 0.015}
\newcommand{\MEDSRavgprobLRMSE}{ 0.108}
\newcommand{\MEDSRavgprobLRBrierref}{ 0.195}
\newcommand{\MEDSRavgprobLRSkillScore}{ 0.447}
\newcommand{\MEDSRavgprobLRACC}{ 0.872}
\newcommand{\MEDSRavgprobLOGRECE}{ 0.134}
\newcommand{\MEDSRavgprobLOGRASCE}{ 0.024}
\newcommand{\MEDSRavgprobLOGRMSE}{ 0.134}
\newcommand{\MEDSRavgprobLOGRBrierref}{ 0.195}
\newcommand{\MEDSRavgprobLOGRSkillScore}{ 0.312}
\newcommand{\MEDSRavgprobLOGRACC}{ 0.866}
\newcommand{\MEDSRavgprobIGHBECE}{ 0.185}
\newcommand{\MEDSRavgprobIGHBASCE}{ 0.050}
\newcommand{\MEDSRavgprobIGHBMSE}{ 0.203}
\newcommand{\MEDSRavgprobIGHBBrierref}{ 0.195}
\newcommand{\MEDSRavgprobIGHBSkillScore}{ -0.039}
\newcommand{\MEDSRavgprobIGHBACC}{ 0.613}
\newcommand{\MEDSRavgprobIGLBECE}{ 0.035}
\newcommand{\MEDSRavgprobIGLBASCE}{ 0.002}
\newcommand{\MEDSRavgprobIGLBMSE}{ 0.102}
\newcommand{\MEDSRavgprobIGLBBrierref}{ 0.195}
\newcommand{\MEDSRavgprobIGLBSkillScore}{ 0.479}
\newcommand{\MEDSRavgprobIGLBACC}{ 0.846}

\newcommand{\MEGPTOSSavgprobUncalibECE}{ 0.268}
\newcommand{\MEGPTOSSavgprobUncalibASCE}{ 0.082}
\newcommand{\MEGPTOSSavgprobUncalibMSE}{ 0.306}
\newcommand{\MEGPTOSSavgprobUncalibBrierref}{ 0.249}
\newcommand{\MEGPTOSSavgprobUncalibSkillScore}{ -0.228}
\newcommand{\MEGPTOSSavgprobUncalibACC}{ 0.474}
\newcommand{\MEGPTOSSavgprobUncalibgASCEcompeasy}{ 0.033}
\newcommand{\MEGPTOSSavgprobUncalibgASCEcompmedium}{ 0.141}
\newcommand{\MEGPTOSSavgprobUncalibgASCEcomphard}{ 0.248}
\newcommand{\MEGPTOSSavgprobUncalibgASCEpromptlenhigh}{ 0.122}
\newcommand{\MEGPTOSSavgprobUncalibgASCElochigh}{ 0.142}
\newcommand{\MEGPTOSSavgprobUncalibgASCElenhigh}{ 0.173}
\newcommand{\MEGPTOSSavgprobUncalibgASCElangC}{ 0.172}
\newcommand{\MEGPTOSSavgprobUncalibgASCElangCPP}{ 0.180}
\newcommand{\MEGPTOSSavgprobUncalibgASCElangCoffeeScript}{ 0.044}
\newcommand{\MEGPTOSSavgprobUncalibgASCElangCommonLisp}{ 0.036}
\newcommand{\MEGPTOSSavgprobUncalibgASCElangDart}{ 0.351}
\newcommand{\MEGPTOSSavgprobUncalibgASCElangElixir}{ 0.074}
\newcommand{\MEGPTOSSavgprobUncalibgASCElangEmacsLisp}{ 0.139}
\newcommand{\MEGPTOSSavgprobUncalibgASCElangFortran}{ 0.187}
\newcommand{\MEGPTOSSavgprobUncalibgASCElangGo}{ 0.096}
\newcommand{\MEGPTOSSavgprobUncalibgASCElangGroovy}{ 0.091}
\newcommand{\MEGPTOSSavgprobUncalibgASCElangHTML}{ 0.223}
\newcommand{\MEGPTOSSavgprobUncalibgASCElangHaskell}{ 0.052}
\newcommand{\MEGPTOSSavgprobUncalibgASCElangJSON}{ 0.017}
\newcommand{\MEGPTOSSavgprobUncalibgASCElangJava}{ 0.247}
\newcommand{\MEGPTOSSavgprobUncalibgASCElangJavaScript}{ 0.139}
\newcommand{\MEGPTOSSavgprobUncalibgASCElangJulia}{ 0.085}
\newcommand{\MEGPTOSSavgprobUncalibgASCElangKotlin}{ 0.027}
\newcommand{\MEGPTOSSavgprobUncalibgASCElangLua}{ 0.135}
\newcommand{\MEGPTOSSavgprobUncalibgASCElangPHP}{ 0.135}
\newcommand{\MEGPTOSSavgprobUncalibgASCElangPascal}{ 0.184}
\newcommand{\MEGPTOSSavgprobUncalibgASCElangPerl}{ 0.069}
\newcommand{\MEGPTOSSavgprobUncalibgASCElangPowerShell}{ 0.064}
\newcommand{\MEGPTOSSavgprobUncalibgASCElangPython}{ 0.026}
\newcommand{\MEGPTOSSavgprobUncalibgASCElangR}{ 0.325}
\newcommand{\MEGPTOSSavgprobUncalibgASCElangRacket}{ 0.116}
\newcommand{\MEGPTOSSavgprobUncalibgASCElangRuby}{ 0.016}
\newcommand{\MEGPTOSSavgprobUncalibgASCElangRust}{ 0.166}
\newcommand{\MEGPTOSSavgprobUncalibgASCElangScala}{ 0.076}
\newcommand{\MEGPTOSSavgprobUncalibgASCElangScheme}{ 0.094}
\newcommand{\MEGPTOSSavgprobUncalibgASCElangShell}{ 0.077}
\newcommand{\MEGPTOSSavgprobUncalibgASCElangSwift}{ 0.029}
\newcommand{\MEGPTOSSavgprobUncalibgASCElangTcl}{ 0.111}
\newcommand{\MEGPTOSSavgprobUncalibgASCElangTypeScript}{ 0.162}
\newcommand{\MEGPTOSSavgprobUncalibgASCElangVimScript}{ 0.097}
\newcommand{\MEGPTOSSavgprobPLATTECE}{ 0.034}
\newcommand{\MEGPTOSSavgprobPLATTASCE}{ 0.002}
\newcommand{\MEGPTOSSavgprobPLATTMSE}{ 0.225}
\newcommand{\MEGPTOSSavgprobPLATTBrierref}{ 0.249}
\newcommand{\MEGPTOSSavgprobPLATTSkillScore}{ 0.097}
\newcommand{\MEGPTOSSavgprobPLATTACC}{ 0.617}
\newcommand{\MEGPTOSSavgprobPLATTgASCEcompeasy}{ 0.013}
\newcommand{\MEGPTOSSavgprobPLATTgASCEcompmedium}{ 0.012}
\newcommand{\MEGPTOSSavgprobPLATTgASCEcomphard}{ 0.038}
\newcommand{\MEGPTOSSavgprobPLATTgASCEpromptlenhigh}{ 0.005}
\newcommand{\MEGPTOSSavgprobPLATTgASCElochigh}{ 0.011}
\newcommand{\MEGPTOSSavgprobPLATTgASCElenhigh}{ 0.012}
\newcommand{\MEGPTOSSavgprobPLATTgASCElangC}{ 0.041}
\newcommand{\MEGPTOSSavgprobPLATTgASCElangCPP}{ 0.050}
\newcommand{\MEGPTOSSavgprobPLATTgASCElangCoffeeScript}{ 0.047}
\newcommand{\MEGPTOSSavgprobPLATTgASCElangCommonLisp}{ 0.062}
\newcommand{\MEGPTOSSavgprobPLATTgASCElangDart}{ 0.092}
\newcommand{\MEGPTOSSavgprobPLATTgASCElangElixir}{ 0.020}
\newcommand{\MEGPTOSSavgprobPLATTgASCElangEmacsLisp}{ 0.017}
\newcommand{\MEGPTOSSavgprobPLATTgASCElangFortran}{ 0.053}
\newcommand{\MEGPTOSSavgprobPLATTgASCElangGo}{ 0.037}
\newcommand{\MEGPTOSSavgprobPLATTgASCElangGroovy}{ 0.023}
\newcommand{\MEGPTOSSavgprobPLATTgASCElangHTML}{ 0.101}
\newcommand{\MEGPTOSSavgprobPLATTgASCElangHaskell}{ 0.033}
\newcommand{\MEGPTOSSavgprobPLATTgASCElangJSON}{ 0.029}
\newcommand{\MEGPTOSSavgprobPLATTgASCElangJava}{ 0.067}
\newcommand{\MEGPTOSSavgprobPLATTgASCElangJavaScript}{ 0.024}
\newcommand{\MEGPTOSSavgprobPLATTgASCElangJulia}{ 0.042}
\newcommand{\MEGPTOSSavgprobPLATTgASCElangKotlin}{ 0.048}
\newcommand{\MEGPTOSSavgprobPLATTgASCElangLua}{ 0.052}
\newcommand{\MEGPTOSSavgprobPLATTgASCElangPHP}{ 0.036}
\newcommand{\MEGPTOSSavgprobPLATTgASCElangPascal}{ 0.023}
\newcommand{\MEGPTOSSavgprobPLATTgASCElangPerl}{ 0.030}
\newcommand{\MEGPTOSSavgprobPLATTgASCElangPowerShell}{ 0.014}
\newcommand{\MEGPTOSSavgprobPLATTgASCElangPython}{ 0.121}
\newcommand{\MEGPTOSSavgprobPLATTgASCElangR}{ 0.136}
\newcommand{\MEGPTOSSavgprobPLATTgASCElangRacket}{ 0.031}
\newcommand{\MEGPTOSSavgprobPLATTgASCElangRuby}{ 0.111}
\newcommand{\MEGPTOSSavgprobPLATTgASCElangRust}{ 0.038}
\newcommand{\MEGPTOSSavgprobPLATTgASCElangScala}{ 0.030}
\newcommand{\MEGPTOSSavgprobPLATTgASCElangScheme}{ 0.027}
\newcommand{\MEGPTOSSavgprobPLATTgASCElangShell}{ 0.017}
\newcommand{\MEGPTOSSavgprobPLATTgASCElangSwift}{ 0.107}
\newcommand{\MEGPTOSSavgprobPLATTgASCElangTcl}{ 0.017}
\newcommand{\MEGPTOSSavgprobPLATTgASCElangTypeScript}{ 0.032}
\newcommand{\MEGPTOSSavgprobPLATTgASCElangVimScript}{ 0.034}
\newcommand{\MEGPTOSSavgprobHBECE}{ 0.025}
\newcommand{\MEGPTOSSavgprobHBASCE}{ 0.001}
\newcommand{\MEGPTOSSavgprobHBMSE}{ 0.226}
\newcommand{\MEGPTOSSavgprobHBBrierref}{ 0.249}
\newcommand{\MEGPTOSSavgprobHBSkillScore}{ 0.094}
\newcommand{\MEGPTOSSavgprobHBACC}{ 0.623}
\newcommand{\MEGPTOSSavgprobHBgASCEcompeasy}{ 0.010}
\newcommand{\MEGPTOSSavgprobHBgASCEcompmedium}{ 0.011}
\newcommand{\MEGPTOSSavgprobHBgASCEcomphard}{ 0.040}
\newcommand{\MEGPTOSSavgprobHBgASCEpromptlenhigh}{ 0.005}
\newcommand{\MEGPTOSSavgprobHBgASCElochigh}{ 0.008}
\newcommand{\MEGPTOSSavgprobHBgASCElenhigh}{ 0.012}
\newcommand{\MEGPTOSSavgprobHBgASCElangC}{ 0.024}
\newcommand{\MEGPTOSSavgprobHBgASCElangCPP}{ 0.025}
\newcommand{\MEGPTOSSavgprobHBgASCElangCoffeeScript}{ 0.016}
\newcommand{\MEGPTOSSavgprobHBgASCElangCommonLisp}{ 0.042}
\newcommand{\MEGPTOSSavgprobHBgASCElangDart}{ 0.093}
\newcommand{\MEGPTOSSavgprobHBgASCElangElixir}{ 0.015}
\newcommand{\MEGPTOSSavgprobHBgASCElangEmacsLisp}{ 0.014}
\newcommand{\MEGPTOSSavgprobHBgASCElangFortran}{ 0.048}
\newcommand{\MEGPTOSSavgprobHBgASCElangGo}{ 0.021}
\newcommand{\MEGPTOSSavgprobHBgASCElangGroovy}{ 0.010}
\newcommand{\MEGPTOSSavgprobHBgASCElangHTML}{ 0.091}
\newcommand{\MEGPTOSSavgprobHBgASCElangHaskell}{ 0.018}
\newcommand{\MEGPTOSSavgprobHBgASCElangJSON}{ 0.023}
\newcommand{\MEGPTOSSavgprobHBgASCElangJava}{ 0.060}
\newcommand{\MEGPTOSSavgprobHBgASCElangJavaScript}{ 0.012}
\newcommand{\MEGPTOSSavgprobHBgASCElangJulia}{ 0.009}
\newcommand{\MEGPTOSSavgprobHBgASCElangKotlin}{ 0.042}
\newcommand{\MEGPTOSSavgprobHBgASCElangLua}{ 0.041}
\newcommand{\MEGPTOSSavgprobHBgASCElangPHP}{ 0.018}
\newcommand{\MEGPTOSSavgprobHBgASCElangPascal}{ 0.015}
\newcommand{\MEGPTOSSavgprobHBgASCElangPerl}{ 0.028}
\newcommand{\MEGPTOSSavgprobHBgASCElangPowerShell}{ 0.008}
\newcommand{\MEGPTOSSavgprobHBgASCElangPython}{ 0.127}
\newcommand{\MEGPTOSSavgprobHBgASCElangR}{ 0.096}
\newcommand{\MEGPTOSSavgprobHBgASCElangRacket}{ 0.018}
\newcommand{\MEGPTOSSavgprobHBgASCElangRuby}{ 0.113}
\newcommand{\MEGPTOSSavgprobHBgASCElangRust}{ 0.026}
\newcommand{\MEGPTOSSavgprobHBgASCElangScala}{ 0.009}
\newcommand{\MEGPTOSSavgprobHBgASCElangScheme}{ 0.003}
\newcommand{\MEGPTOSSavgprobHBgASCElangShell}{ 0.009}
\newcommand{\MEGPTOSSavgprobHBgASCElangSwift}{ 0.102}
\newcommand{\MEGPTOSSavgprobHBgASCElangTcl}{ 0.005}
\newcommand{\MEGPTOSSavgprobHBgASCElangTypeScript}{ 0.021}
\newcommand{\MEGPTOSSavgprobHBgASCElangVimScript}{ 0.034}
\newcommand{\MEGPTOSSavgprobLRECE}{ 0.041}
\newcommand{\MEGPTOSSavgprobLRASCE}{ 0.002}
\newcommand{\MEGPTOSSavgprobLRMSE}{ 0.187}
\newcommand{\MEGPTOSSavgprobLRBrierref}{ 0.249}
\newcommand{\MEGPTOSSavgprobLRSkillScore}{ 0.251}
\newcommand{\MEGPTOSSavgprobLRACC}{ 0.725}
\newcommand{\MEGPTOSSavgprobLRgASCEcompeasy}{ 0.005}
\newcommand{\MEGPTOSSavgprobLRgASCEcompmedium}{ 0.004}
\newcommand{\MEGPTOSSavgprobLRgASCEcomphard}{ 0.004}
\newcommand{\MEGPTOSSavgprobLRgASCEpromptlenhigh}{ 0.003}
\newcommand{\MEGPTOSSavgprobLRgASCElochigh}{ 0.004}
\newcommand{\MEGPTOSSavgprobLRgASCElenhigh}{ 0.003}
\newcommand{\MEGPTOSSavgprobLRgASCElangC}{ 0.022}
\newcommand{\MEGPTOSSavgprobLRgASCElangCPP}{ 0.032}
\newcommand{\MEGPTOSSavgprobLRgASCElangCoffeeScript}{ 0.039}
\newcommand{\MEGPTOSSavgprobLRgASCElangCommonLisp}{ 0.031}
\newcommand{\MEGPTOSSavgprobLRgASCElangDart}{ 0.018}
\newcommand{\MEGPTOSSavgprobLRgASCElangElixir}{ 0.015}
\newcommand{\MEGPTOSSavgprobLRgASCElangEmacsLisp}{ 0.037}
\newcommand{\MEGPTOSSavgprobLRgASCElangFortran}{ 0.028}
\newcommand{\MEGPTOSSavgprobLRgASCElangGo}{ 0.024}
\newcommand{\MEGPTOSSavgprobLRgASCElangGroovy}{ 0.028}
\newcommand{\MEGPTOSSavgprobLRgASCElangHTML}{ 0.015}
\newcommand{\MEGPTOSSavgprobLRgASCElangHaskell}{ 0.026}
\newcommand{\MEGPTOSSavgprobLRgASCElangJSON}{ 0.022}
\newcommand{\MEGPTOSSavgprobLRgASCElangJava}{ 0.018}
\newcommand{\MEGPTOSSavgprobLRgASCElangJavaScript}{ 0.024}
\newcommand{\MEGPTOSSavgprobLRgASCElangJulia}{ 0.035}
\newcommand{\MEGPTOSSavgprobLRgASCElangKotlin}{ 0.019}
\newcommand{\MEGPTOSSavgprobLRgASCElangLua}{ 0.033}
\newcommand{\MEGPTOSSavgprobLRgASCElangPHP}{ 0.020}
\newcommand{\MEGPTOSSavgprobLRgASCElangPascal}{ 0.013}
\newcommand{\MEGPTOSSavgprobLRgASCElangPerl}{ 0.041}
\newcommand{\MEGPTOSSavgprobLRgASCElangPowerShell}{ 0.022}
\newcommand{\MEGPTOSSavgprobLRgASCElangPython}{ 0.082}
\newcommand{\MEGPTOSSavgprobLRgASCElangR}{ 0.049}
\newcommand{\MEGPTOSSavgprobLRgASCElangRacket}{ 0.049}
\newcommand{\MEGPTOSSavgprobLRgASCElangRuby}{ 0.046}
\newcommand{\MEGPTOSSavgprobLRgASCElangRust}{ 0.027}
\newcommand{\MEGPTOSSavgprobLRgASCElangScala}{ 0.038}
\newcommand{\MEGPTOSSavgprobLRgASCElangScheme}{ 0.036}
\newcommand{\MEGPTOSSavgprobLRgASCElangShell}{ 0.013}
\newcommand{\MEGPTOSSavgprobLRgASCElangSwift}{ 0.030}
\newcommand{\MEGPTOSSavgprobLRgASCElangTcl}{ 0.021}
\newcommand{\MEGPTOSSavgprobLRgASCElangTypeScript}{ 0.023}
\newcommand{\MEGPTOSSavgprobLRgASCElangVimScript}{ 0.053}
\newcommand{\MEGPTOSSavgprobLOGRECE}{ 0.272}
\newcommand{\MEGPTOSSavgprobLOGRASCE}{ 0.074}
\newcommand{\MEGPTOSSavgprobLOGRMSE}{ 0.272}
\newcommand{\MEGPTOSSavgprobLOGRBrierref}{ 0.249}
\newcommand{\MEGPTOSSavgprobLOGRSkillScore}{ -0.092}
\newcommand{\MEGPTOSSavgprobLOGRACC}{ 0.728}
\newcommand{\MEGPTOSSavgprobLOGRgASCEcompeasy}{ 0.080}
\newcommand{\MEGPTOSSavgprobLOGRgASCEcompmedium}{ 0.085}
\newcommand{\MEGPTOSSavgprobLOGRgASCEcomphard}{ 0.056}
\newcommand{\MEGPTOSSavgprobLOGRgASCEpromptlenhigh}{ 0.072}
\newcommand{\MEGPTOSSavgprobLOGRgASCElochigh}{ 0.086}
\newcommand{\MEGPTOSSavgprobLOGRgASCElenhigh}{ 0.076}
\newcommand{\MEGPTOSSavgprobLOGRgASCElangC}{ 0.056}
\newcommand{\MEGPTOSSavgprobLOGRgASCElangCPP}{ 0.021}
\newcommand{\MEGPTOSSavgprobLOGRgASCElangCoffeeScript}{ 0.141}
\newcommand{\MEGPTOSSavgprobLOGRgASCElangCommonLisp}{ 0.057}
\newcommand{\MEGPTOSSavgprobLOGRgASCElangDart}{ 0.014}
\newcommand{\MEGPTOSSavgprobLOGRgASCElangElixir}{ 0.065}
\newcommand{\MEGPTOSSavgprobLOGRgASCElangEmacsLisp}{ 0.091}
\newcommand{\MEGPTOSSavgprobLOGRgASCElangFortran}{ 0.168}
\newcommand{\MEGPTOSSavgprobLOGRgASCElangGo}{ 0.030}
\newcommand{\MEGPTOSSavgprobLOGRgASCElangGroovy}{ 0.161}
\newcommand{\MEGPTOSSavgprobLOGRgASCElangHTML}{ 0.113}
\newcommand{\MEGPTOSSavgprobLOGRgASCElangHaskell}{ 0.122}
\newcommand{\MEGPTOSSavgprobLOGRgASCElangJSON}{ 0.061}
\newcommand{\MEGPTOSSavgprobLOGRgASCElangJava}{ 0.094}
\newcommand{\MEGPTOSSavgprobLOGRgASCElangJavaScript}{ 0.045}
\newcommand{\MEGPTOSSavgprobLOGRgASCElangJulia}{ 0.111}
\newcommand{\MEGPTOSSavgprobLOGRgASCElangKotlin}{ 0.057}
\newcommand{\MEGPTOSSavgprobLOGRgASCElangLua}{ 0.030}
\newcommand{\MEGPTOSSavgprobLOGRgASCElangPHP}{ 0.066}
\newcommand{\MEGPTOSSavgprobLOGRgASCElangPascal}{ 0.055}
\newcommand{\MEGPTOSSavgprobLOGRgASCElangPerl}{ 0.051}
\newcommand{\MEGPTOSSavgprobLOGRgASCElangPowerShell}{ 0.083}
\newcommand{\MEGPTOSSavgprobLOGRgASCElangPython}{ 0.178}
\newcommand{\MEGPTOSSavgprobLOGRgASCElangR}{ 0.049}
\newcommand{\MEGPTOSSavgprobLOGRgASCElangRacket}{ 0.064}
\newcommand{\MEGPTOSSavgprobLOGRgASCElangRuby}{ 0.112}
\newcommand{\MEGPTOSSavgprobLOGRgASCElangRust}{ 0.133}
\newcommand{\MEGPTOSSavgprobLOGRgASCElangScala}{ 0.119}
\newcommand{\MEGPTOSSavgprobLOGRgASCElangScheme}{ 0.072}
\newcommand{\MEGPTOSSavgprobLOGRgASCElangShell}{ 0.091}
\newcommand{\MEGPTOSSavgprobLOGRgASCElangSwift}{ 0.034}
\newcommand{\MEGPTOSSavgprobLOGRgASCElangTcl}{ 0.085}
\newcommand{\MEGPTOSSavgprobLOGRgASCElangTypeScript}{ 0.047}
\newcommand{\MEGPTOSSavgprobLOGRgASCElangVimScript}{ 0.174}
\newcommand{\MEGPTOSSavgprobIGHBECE}{ 0.164}
\newcommand{\MEGPTOSSavgprobIGHBASCE}{ 0.041}
\newcommand{\MEGPTOSSavgprobIGHBMSE}{ 0.259}
\newcommand{\MEGPTOSSavgprobIGHBBrierref}{ 0.249}
\newcommand{\MEGPTOSSavgprobIGHBSkillScore}{ -0.038}
\newcommand{\MEGPTOSSavgprobIGHBACC}{ 0.589}
\newcommand{\MEGPTOSSavgprobIGHBgASCEcompeasy}{ 0.016}
\newcommand{\MEGPTOSSavgprobIGHBgASCEcompmedium}{ 0.077}
\newcommand{\MEGPTOSSavgprobIGHBgASCEcomphard}{ 0.147}
\newcommand{\MEGPTOSSavgprobIGHBgASCEpromptlenhigh}{ 0.068}
\newcommand{\MEGPTOSSavgprobIGHBgASCElochigh}{ 0.075}
\newcommand{\MEGPTOSSavgprobIGHBgASCElenhigh}{ 0.079}
\newcommand{\MEGPTOSSavgprobIGHBgASCElangC}{ 0.093}
\newcommand{\MEGPTOSSavgprobIGHBgASCElangCPP}{ 0.099}
\newcommand{\MEGPTOSSavgprobIGHBgASCElangCoffeeScript}{ 0.029}
\newcommand{\MEGPTOSSavgprobIGHBgASCElangCommonLisp}{ 0.033}
\newcommand{\MEGPTOSSavgprobIGHBgASCElangDart}{ 0.221}
\newcommand{\MEGPTOSSavgprobIGHBgASCElangElixir}{ 0.036}
\newcommand{\MEGPTOSSavgprobIGHBgASCElangEmacsLisp}{ 0.056}
\newcommand{\MEGPTOSSavgprobIGHBgASCElangFortran}{ 0.127}
\newcommand{\MEGPTOSSavgprobIGHBgASCElangGo}{ 0.063}
\newcommand{\MEGPTOSSavgprobIGHBgASCElangGroovy}{ 0.069}
\newcommand{\MEGPTOSSavgprobIGHBgASCElangHTML}{ 0.198}
\newcommand{\MEGPTOSSavgprobIGHBgASCElangHaskell}{ 0.026}
\newcommand{\MEGPTOSSavgprobIGHBgASCElangJSON}{ 0.014}
\newcommand{\MEGPTOSSavgprobIGHBgASCElangJava}{ 0.194}
\newcommand{\MEGPTOSSavgprobIGHBgASCElangJavaScript}{ 0.094}
\newcommand{\MEGPTOSSavgprobIGHBgASCElangJulia}{ 0.060}
\newcommand{\MEGPTOSSavgprobIGHBgASCElangKotlin}{ 0.020}
\newcommand{\MEGPTOSSavgprobIGHBgASCElangLua}{ 0.104}
\newcommand{\MEGPTOSSavgprobIGHBgASCElangPHP}{ 0.070}
\newcommand{\MEGPTOSSavgprobIGHBgASCElangPascal}{ 0.111}
\newcommand{\MEGPTOSSavgprobIGHBgASCElangPerl}{ 0.051}
\newcommand{\MEGPTOSSavgprobIGHBgASCElangPowerShell}{ 0.020}
\newcommand{\MEGPTOSSavgprobIGHBgASCElangPython}{ 0.071}
\newcommand{\MEGPTOSSavgprobIGHBgASCElangR}{ 0.289}
\newcommand{\MEGPTOSSavgprobIGHBgASCElangRacket}{ 0.067}
\newcommand{\MEGPTOSSavgprobIGHBgASCElangRuby}{ 0.055}
\newcommand{\MEGPTOSSavgprobIGHBgASCElangRust}{ 0.100}
\newcommand{\MEGPTOSSavgprobIGHBgASCElangScala}{ 0.069}
\newcommand{\MEGPTOSSavgprobIGHBgASCElangScheme}{ 0.034}
\newcommand{\MEGPTOSSavgprobIGHBgASCElangShell}{ 0.039}
\newcommand{\MEGPTOSSavgprobIGHBgASCElangSwift}{ 0.045}
\newcommand{\MEGPTOSSavgprobIGHBgASCElangTcl}{ 0.041}
\newcommand{\MEGPTOSSavgprobIGHBgASCElangTypeScript}{ 0.088}
\newcommand{\MEGPTOSSavgprobIGHBgASCElangVimScript}{ 0.098}
\newcommand{\MEGPTOSSavgprobIGLBECE}{ 0.022}
\newcommand{\MEGPTOSSavgprobIGLBASCE}{ 0.001}
\newcommand{\MEGPTOSSavgprobIGLBMSE}{ 0.185}
\newcommand{\MEGPTOSSavgprobIGLBBrierref}{ 0.249}
\newcommand{\MEGPTOSSavgprobIGLBSkillScore}{ 0.258}
\newcommand{\MEGPTOSSavgprobIGLBACC}{ 0.724}
\newcommand{\MEGPTOSSavgprobIGLBgASCEcompeasy}{ 0.002}
\newcommand{\MEGPTOSSavgprobIGLBgASCEcompmedium}{ 0.004}
\newcommand{\MEGPTOSSavgprobIGLBgASCEcomphard}{ 0.007}
\newcommand{\MEGPTOSSavgprobIGLBgASCEpromptlenhigh}{ 0.002}
\newcommand{\MEGPTOSSavgprobIGLBgASCElochigh}{ 0.001}
\newcommand{\MEGPTOSSavgprobIGLBgASCElenhigh}{ 0.002}
\newcommand{\MEGPTOSSavgprobIGLBgASCElangC}{ 0.021}
\newcommand{\MEGPTOSSavgprobIGLBgASCElangCPP}{ 0.037}
\newcommand{\MEGPTOSSavgprobIGLBgASCElangCoffeeScript}{ 0.056}
\newcommand{\MEGPTOSSavgprobIGLBgASCElangCommonLisp}{ 0.042}
\newcommand{\MEGPTOSSavgprobIGLBgASCElangDart}{ 0.015}
\newcommand{\MEGPTOSSavgprobIGLBgASCElangElixir}{ 0.022}
\newcommand{\MEGPTOSSavgprobIGLBgASCElangEmacsLisp}{ 0.033}
\newcommand{\MEGPTOSSavgprobIGLBgASCElangFortran}{ 0.038}
\newcommand{\MEGPTOSSavgprobIGLBgASCElangGo}{ 0.020}
\newcommand{\MEGPTOSSavgprobIGLBgASCElangGroovy}{ 0.031}
\newcommand{\MEGPTOSSavgprobIGLBgASCElangHTML}{ 0.051}
\newcommand{\MEGPTOSSavgprobIGLBgASCElangHaskell}{ 0.019}
\newcommand{\MEGPTOSSavgprobIGLBgASCElangJSON}{ 0.034}
\newcommand{\MEGPTOSSavgprobIGLBgASCElangJava}{ 0.003}
\newcommand{\MEGPTOSSavgprobIGLBgASCElangJavaScript}{ 0.026}
\newcommand{\MEGPTOSSavgprobIGLBgASCElangJulia}{ 0.044}
\newcommand{\MEGPTOSSavgprobIGLBgASCElangKotlin}{ 0.032}
\newcommand{\MEGPTOSSavgprobIGLBgASCElangLua}{ 0.028}
\newcommand{\MEGPTOSSavgprobIGLBgASCElangPHP}{ 0.016}
\newcommand{\MEGPTOSSavgprobIGLBgASCElangPascal}{ 0.037}
\newcommand{\MEGPTOSSavgprobIGLBgASCElangPerl}{ 0.034}
\newcommand{\MEGPTOSSavgprobIGLBgASCElangPowerShell}{ 0.025}
\newcommand{\MEGPTOSSavgprobIGLBgASCElangPython}{ 0.030}
\newcommand{\MEGPTOSSavgprobIGLBgASCElangR}{ 0.006}
\newcommand{\MEGPTOSSavgprobIGLBgASCElangRacket}{ 0.035}
\newcommand{\MEGPTOSSavgprobIGLBgASCElangRuby}{ 0.004}
\newcommand{\MEGPTOSSavgprobIGLBgASCElangRust}{ 0.011}
\newcommand{\MEGPTOSSavgprobIGLBgASCElangScala}{ 0.019}
\newcommand{\MEGPTOSSavgprobIGLBgASCElangScheme}{ 0.018}
\newcommand{\MEGPTOSSavgprobIGLBgASCElangShell}{ 0.017}
\newcommand{\MEGPTOSSavgprobIGLBgASCElangSwift}{ 0.045}
\newcommand{\MEGPTOSSavgprobIGLBgASCElangTcl}{ 0.029}
\newcommand{\MEGPTOSSavgprobIGLBgASCElangTypeScript}{ 0.021}
\newcommand{\MEGPTOSSavgprobIGLBgASCElangVimScript}{ 0.043}
\title{
Multicalibration for LLM-based Code Generation
}

\author{Viola Campos}
\email{viola.campos@hs-rm.de}
\orcid{0009-0004-0188-6154}

\affiliation{%
  \institution{RheinMain University of Applied Sciences}
  \city{Wiesbaden}
  \country{Germany}
}

\author{Robin Kuschnereit}
\email{robin.kuschnereit@gmail.com}
\affiliation{%
  \institution{RheinMain University of Applied Sciences}
  \city{Wiesbaden}
  \country{Germany}
}

\author{Adrian Ulges}
\email{adrian.ulges@hs-rm.de}
\orcid{0009-0001-1915-2464}
\affiliation{%
  \institution{RheinMain University of Applied Sciences}
  \city{Wiesbaden}
  \country{Germany}
}

\renewcommand{\shortauthors}{Campos et al.}

\begin{abstract}
As AI-based code generation becomes widespread, researchers are investigating the calibration of code LLMs — ensuring their confidence scores faithfully represent the true likelihood of code correctness.
To do so, we investigate {\it multicalibration}, which can capture additional factors about a coding problem, such as complexity, code length, or programming language used. We study four multicalibration approaches on three function synthesis benchmarks, using latest-generation code LLMs (Qwen3 Coder, GPT-OSS, DeepSeek-R1-Distill). Our results demonstrate that multicalibration can yield distinct improvements over both uncalibrated token likelihoods ($+ 1.03$ in skill score) and baseline calibrations ($+ 0.37$ in skill score).  We study the influence of the aforementioned factors in ablations, and make our dataset (consisting of code generations, likelihoods, and correctness labels) available for future research on code LLM calibration.
\end{abstract}



\maketitle
\pagestyle{plain}

\section{Introduction}

AI-based code generators are fundamentally transforming software development practice: By prompting Large Language Models (LLMs), code can be produced and modified much faster and easier, leading to substantial efficiency gains and reducing the knowledge level required by programmers~\cite{cui2024effects}. 
However, this promise of higher efficiency comes with the risk of degrading code quality, as LLMs (being probabilistic models) are inherently unreliable~\cite{ChenLD24deep_dive, TambonDNKDA25} and developers get overwhelmed as more and more code is generated. 
Therefore, approaches are required to support developers with reviewing code efficiently. 

Two such approaches are {\it uncertainty estimation} and {\it calibration}. Both are closely related: Uncertainty estimation (UE) quantifies the model's confidence level regarding the correctness of its prediction,
and calibration corrects the model's confidence scores to better reflect the expected correctness. 
Applied to code LLMs, such approaches could pinpoint uncertain predictions, which would improve the efficiency of code reviewing and reduce its mental load. 

Uncertainty estimation and calibration of LLMs have been studied 
mostly for {\it natural language understanding (NLU)} tasks, 
where hallucinations pose a well-known problem~\cite{liu25survey}. 
There are only few works 
for {\it code LLMs}, particularly when it comes to 
post-hoc calibration~\cite{Spiess0PPRAJDA25}.

In this paper, we argue that a successful calibration of code LLMs should take additional code-related information into account. This could come in various forms, such as complexity measures of the code or its specification, or the programming language used.
Though this information could easily be collected and leveraged in a calibration, 
we are not aware of any studies on the subject. 
We fill this gap by using {\it multicalibration}~\cite{HebertJohnsonK18}, which groups inputs according to semantic facets and applies a group-specific calibration, and has been applied for hallucination detection in question answering before ~\cite{DetommasoLF024}. We apply multi-calibration to code generation by grouping coding problems by complexity and/or programming languague. Thereby, our contributions are the following:
\begin{itemize}
    \item To the best of our knowledge, our study is the first to investigate multicalibration for code generation. We investigate four multicalibration approaches, using groups based on complexity, code length, prompt length, and the programming language used. 
    \item We conduct an evaluation 
    using latest open-weight reasoning models (namely 
    \emph{Qwen3-Coder-30B-A3B-Instruct}~\cite{yang25qwen3}, \emph{OpenAI GPT-OSS-20B}~\cite{openai2025gptoss120bgptoss20bmodel} and \emph{DeepSeek-R1-Distill-Qwen-32B}~\cite{guo25deepseek}) on three code generation datasets: the multilingual benchmarks MultiPL-E~\cite{CassanoGNNPPYZAFGGJ23} and McEval~\cite{ChaiL0YJLS0RGWW25}, which offer $20$ and $40$ programming languages, respectively, and LiveCodeBench~\cite{JainHGLYZWSSS25}, which -- since very recent -- minimizes the risk of data leakage. Our results show that calibration benefits strongly from taking the aforementioned code factors into account, with accuracy on distinguishing correct from incorrect solutions improving from $+13.2\%$ to $+58.3\%$.  
    \item To foster future research on this important issue, we make our Dataset \emph{CALIBRI} available as a contribution\footnote{\url{https://huggingface.co/datasets/lavis-nlp/CALIBRI}}. \emph{CALIBRI} includes correctness labels and token likelihoods for latest-generation LLMs on a fresh dataset with relatively low risk of data leakage.
\end{itemize}



%


\section{Related Work}
\label{sec:relatedwork}
Due to their large-scale pretraining and their enormous capacity, LLMs for code have demonstrated impressive problem-solving skills over a wide range of programming tasks~\cite{zhao_survey_2023, HouZLYWLLLGW24}. 
However, their accuracy remains limited, as results in program repair demonstrate~\cite{JimenezYWYPPN24,campos25empiricalapr}. Moreover, 
LLMs have been shown to occasionally produce defective code~\cite{ChenLD24deep_dive, TambonDNKDA25}, 
e.g., by reproducing known bugs~\cite{JesseADM23} or generating potentially risky vulnerabilities~\cite{PearceATDK25}. 
These limitations pose a significant threat to the reliability of code LLMs, 
hence quality control of code generation is a vital issue.

\paragraph{UE and Calibration for Large Language Models} 
Uncertainty estimation and model calibration are long-standing topics 
in probabilistic modeling and classification~\cite{glenn1950verification, steyerberg2010assessing}. Recent work has extended these concepts to LLMs, although research has focused on natural language tasks rather than generative settings~\cite{geng-etal-2024-survey}. Approaches for quantifying uncertainty in generative LLMs can be broadly categorized into black-box and white-box approaches~\cite{Shorinwa25UQsurvey}: Black-box methods operate on input–output pairs only, e.g., \emph{verbalized uncertainty}, where the model is prompted to express confidence in its output~\cite{TianMZSRYFM23, Kadavath22, lin22teachingtoexpress}, and \emph{consistency-based methods}, that estimate uncertainty from agreement across multiple generations~\cite{QiuM24, LinT024, manakul23selfcheckgpt}. White-box methods exploit internal model signals: \emph{information-theoretic approaches} derive confidence scores from output probabilities~\cite{FomichevaSYBGFA20, DuanC0ZWXKX24}, while \emph{introspective methods} analyze LLMs internal activations, specifically, self-attention scores
and hidden states to estimate uncertainty~\cite{snyder24earlydetection, bui2025correctness}.

\paragraph{Post-hoc Calibration} 
Post-hoc scaling techniques such as Platt or temperature scaling~\cite{platt1999probabilistic} or Histogram Binning~\cite{zadrozny01histogrambinning} are commonly used to improve predictive calibration. Originally motivated by fairness and acountability, Hébert-Johnson et al. introduced \emph{multiaccuracy} and \emph{multicalibration}~\cite{HebertJohnsonK18}, extending standard calibration to multiple, possibly intersecting groups. Recent work has started adapting multicalibration to LLMs: Detommaso et al. apply multicalibration algorithms to enhance the reliability of confidence scores in LLM outputs for question-answering tasks~\cite{DetommasoLF024}, and Hansen et al. evaluate multicalibration methods across a broad range of classification models on tabular, image and language datasets, finding benefits in many settings, especially for large vision and language models~\cite{HansenDNS24}.

\paragraph{Uncertainty Estimation and Calibration for Code Generation}




All the above works have been conducted on natural language tasks. In contrast, research on \h{uncertainty estimation in {\it code generation}} is scarcer and has concentrated on two areas. First, several works fine-tune specialized LLMs for correctness estimation~\cite{zhou23codebertscore,hossain2024,reddy2025swerank};
these methods risk overfitting to their limited supervised bug training sets, such as defects4j~\cite{JustJE14defects4j} or Swe-Bench-Lite~\cite{Jimenez24SWEBench}.  
Second, reflective or judge-LLM approaches run additional reasoning steps to assess (or improve) their outputs~\cite{tong24codejudge,wang2025mctsjudge}. 
When it comes to {\it post-hoc confidence estimation / calibration} of off-the-shelf code LLMs, Zhou et al.~\cite{ZhouS024} show for four classification-based code understanding tasks, that vanilla code LLMs are not well-calibrated per se. Spiess et al.~\cite{Spiess0PPRAJDA25} show similar results for generative code LLMs and compare Platt-scaling as a standard calibration approach with self-reflection.  Beyond these, multicalibration remains unstudied for code LLMs, and a rigorous comparisons of all of the strategies mentioned in Section~\ref{sec:relatedwork} is lacking. 
A recent survey confirmed that the field of LLM-based code correctness estimation remains in its early stages~\cite{he2025courtroomsurvey}.

\section{Background}
In this section, we introduce the concepts of calibration and multicalibration for code generation, and describe common metrics for assessing calibration quality.
We assume the input prompt $X$ to consist of a task description and/or code context, and the output  $Z = z_{1:n}$ to be a sequence of code tokens.
$Z$'s correctness can be assessed 
(e.g. through unit tests), resulting in a correctness label $Y {\in} \{0,1\}$ with $Y{=}1$ iff the predicted code $Z$ is correct.
Also, we define a {\it confidence score} $\hat{p}$ representing the model's estimated confidence in the predicted code to be correct. 
$\hat{p}$ is derived from the model's token likelihoods $P(z_i | z_{<i},X)$ for the individual code tokens on the sequence using different aggregation strategies as described in Section~\ref{sec:initial_confidence}.

\subsection{Calibrated Confidences} 
\label{sec:calibration}

Calibration 
aims at measuring how accurately a model's predicted confidence in a certain outcome depicts the actual frequency of that outcome. For example, we assume a weather prediction model to be well calibrated if, on 90 out of 100 days on which the model predicted $90\%$ confidence of rain the next day, it actually rained.
Formally, a model is considered perfectly {\it  calibrated}~\cite{GuoPSW17,JiangADN21} if its confidence $\hat{p}$ matches the true  probability of correctness:
\begin{equation}    \label{eq:perfect_calibration}
 P( Y{=}1 \mid 
 \hat{p} {=} p) = p  \quad \forall p \in[0,1]
\end{equation}
In other words, perfect calibration means that the model's predicted confidence $\hat{p}$ matches the empirical accuracy, i.e. the fraction of correct predictions among all outputs with confidence $\hat{p}$.
As discussed by Guo et al,~\cite{GuoPSW17}, the probability in Equation~\ref{eq:perfect_calibration} cannot be computed exactly with a finite number of samples, since  $ \hat{p}$ is a continuous random variable. Therefore, calibration in practice is conducted by  partitioning confidences into $M$ disjoint bins of equal width:
\begin{equation*}
    \left[ \frac{1}{M}\right] = \left \{ \frac{i}{M} \right \}^M_{i=1}
\end{equation*}
Then, given a set of samples $(x_1,y_1,z_1,\hat{p}_1), ..., (x_n,y_n,z_n,\hat{p}_n)$, each with correctness label $y_i$ and confidence $\hat{p}_i$, we assign the confidences $\hat{p}_1,...,\hat{p}_n$ to their respective bins, obtaining bin-wise sets of confidences ($S_m$) and indices ($B_m$) for $m{=}1,...M$:
\begin{align}
\label{eq:bins}
S_m &:= \begin{cases}
\left\{ \hat{p}_i \;\Big|\; \frac{m-1}{M} \le \hat{p}_i < \frac{m}{M},\; i \in \{1,\ldots,n\} \right\} &\text{if } m < M \\[0.5em]
\left\{ \hat{p}_i \;\Big|\; \frac{M-1}{M} \le \hat{p}_i \le 1,\; i \in \{1,\ldots,n\} \right\} & \text{if } m = M
\end{cases}\\
B_m &:= \left\{ i \in \{1,\ldots,n\} \;\Big|\; \hat{p}_i \in S_m \right\}
\end{align}
A model is then considered calibrated if, \emph{within each bin}, the mean predicted confidence equals the empirical fraction of correct predictions. 


For each bin $B_m$, we aggregate the model's accuracy and confidence as: 
\begin{align}
    \text{acc}(B_m) &= \frac{1}{|B_m|}\sum_{i \in B_m} \mathbf{1}(y_i=1) \\
    \text{conf}(B_m) &=  \frac{1}{|B_m|}\sum_{i \in B_m} \hat{p}_i
\end{align}
A perfectly calibrated model will have $\text{acc}(B_m) = \text{conf}(B_m)$ for all $m =1,...,M$. 
This relationship can be visualized in a \emph{reliability diagram}, which plots the empirical accuracy $\text{acc}(B_m)$ against the mean confidence $\text{conf}(B_m)$ for each bin, as shown in Figure~\ref{fig:bar_charts}. 

\subsubsection{Calibration Measures} 
While reliability diagrams provide an intuitive visualization, a scalar summary measure of (mis)calibration is often preferred. Therefore, several measures are commonly used to quantify the calibration error of a model: 

The \emph{Expected Calibration Error (ECE)~\cite{NaeiniCH15}}  approximates the expected difference between accuracy and confidence as a weighted average of the bins' deviations between accuracy and confidence:


\begin{equation}
    \text{ECE} := \sum_{m=1}^M \frac{|B_m|}{n} \cdot \left | \text{acc}(B_m) - \text{conf}(B_m)\right | 
\end{equation}
where $n$ is the total number of samples.

\paragraph{Brier Score~\cite{glenn1950verification}} \emph{ECE} is intuitive but can be misleading: a naive predictor, whose confidence is always the mean accuracy across all samples $p_r := 1/n \cdot \sum_{i=1}^n y_i$ (also called the {\it base rate}) could yield a deceptively low \emph{ECE}, since all predictions fall into a single bin. An alternative measure avoiding this weakness is the Brier score $\mathcal{B}$, which quantifies the mean squared difference between confidence and true correctness label:
\begin{equation}
    \mathcal{B} := \frac{1}{n}\sum_{i=1}^n(\hat{p}_i - y_i )^2
\end{equation}
A \emph{Brier score} of $0$ implies perfect calibration, with confidences $\hat{p}_i {=} 1$  for correct outputs and $\hat{p}_i {=} 0$ for incorrect ones. Random binary confidences are expected to yield $\mathcal{B}=0.25$ for a balanced dataset ($50\%$ correct and $50\%$ incorrect), while $\mathcal{B}=1$ indicates complete confidence in the wrong outcome.  

\paragraph{Brier Skill Score (BSS)~\cite{glenn1950verification}} For a more interpretable comparison, particularly in unbalanced scenarios, we also report the \emph{Brier Skill Score}: Consider a naive, so-called 
\emph{unskilled predictor}, which simply assigns every sample the base-rate confidence $p_r$. This baseline's Brier Score can be derived in closed form as:
\begin{equation}
    \mathcal{B}_{ref} = p_r \cdot (1-p_r), 
\end{equation}
and the Brier Skill score measures the improvement over this baseline:
\begin{equation}
    \text{BSS} := \frac{ \mathcal{B}_{ref} -  \mathcal{B}}{ \mathcal{B}_{ref}}
\end{equation}
where positive $\text{BSS}$ values show improvement over the unskilled baseline, and negative values show a deterioration.
When $p_r \in \{0, 1\}$, the denominator is zero; by convention, we set $\text{BSS} = 1$ if $\mathcal{B} = 0$ (perfect prediction) and $\text{BSS} = -\infty$ otherwise.

\subsection{Multicalibrated Confidences}
\label{sec:multicalibration}

Following Detommaso et al.~\cite{DetommasoLF024}, we argue that the guarantee provided by standard calibration (Equation \ref{eq:perfect_calibration}) is insufficient for generative language model applications, since LLM performance varies widely across different types of tasks. Consider, for example, two prompt-completion pairs, respectively asking a Code LLM to implement either a bubble sort function or a fully homomorphic encryption scheme. The probability of correctness should differ substantially between these cases, yet calibration is a \emph{marginal} guarantee that may average over both cases. As a result, a calibrated model might produce confidence scores that are systematically overconfident on complex tasks and underconfident on simple tasks, which is not in conflict with even perfect calibration. Ideally, confidence estimates should be calibrated conditionally on as much fine-grained information as possible about the prompt. \emph{Multicalibration}, introduced by Hébert-Johnson et al.~\cite{HebertJohnsonK18}, provides exactly this type of conditional guarantee, ensuring that predicted probabilities align with empirical accuracy across multiple relevant subgroups of  inputs. 

\subsubsection{Groups}
\label{sec:groups}
Multicalibration introduces the notion of  \emph{groups}, i.e. sets of prompt-completion pairs sharing a common property, such as a level of code complexity or a programming language. Let $\mathcal{D}$ be the distribution of prompt-completion pairs $(X,Z) \sim \mathcal{D}$. A \emph{group function} $g : \mathcal{D} \to \{0,1\}$ indicates group membership, i.e. $g(x, z) = 1$ iff a pair $(x,z)$ belongs to Group $g$. Analogous to the confidence bins $B_m$, we define for each group function $g$ the corresponding index set as
\begin{equation}
    G := \left\{ i \in\{1,...,n\} \mid g(x_i, z_i) =1\right\}
\end{equation}
A set of groups $\mathcal{G}$ is therefore identified by a set of indicator functions. Note that prompt-completion pairs can have multiple non-mutually-exclusive attributes, so that groups in $\mathcal{G}$ may intersect.

\paragraph{Group-conditional unbiasedness (multiaccuracy)}
Before defining multicalibration, we consider a weaker notion: \emph{group-conditional unbiasedness} (also known as \emph{multiaccuracy})~\cite{HebertJohnsonK18,KimGZ19}. 
A scoring function $\hat{p}$ is \emph{group-conditionally unbiased} with respect to a data distribution $\mathcal{D}$ and a set of groups $\mathcal{G}$ if 
\begin{equation}
\label{eq:multiaccuracy}
    \mathbb{E}_\mathcal{D}[Y-\hat{p} \mid g(X)=1] = 0, \quad \forall g \in \mathcal{G}.
\end{equation}
This condition ensures that the mean predicted confidence equals the true accuracy within each group, but it does not yet account for distributional calibration errors across confidence levels.

\paragraph{Group Average Squared Calibration Error (gASCE)} To quantify calibration error conditioned jointly on the groups $\mathcal{G}$ and a set of disjoint confidence bins $B_1,...,B_M$, we use the \emph{group average squared calibration error}~\cite{DetommasoLF024}.

For a single group $g$, the gASCE is defined as
\begin{align}
    gASCE(g) & := \sum_{m=1}^M P(m|g)  \cdot (\Delta_{g,m})^2, \\
    \text{where}\quad \Delta_{g,m} &:= \frac{1}{|G \cap B_m |} \cdot \sum_{j \in G \cap B_m } (y_j - \hat{p}_j) \nonumber \\
    \text{and} \quad  P(m|g) &= \frac{|G \cap B_m|}{|G|} \nonumber
\end{align}



\subsubsection{Multicalibration}
Recall that classical calibration as defined in Section \ref{sec:calibration} requires a model to be unbiased conditional on a set of disjoint confidence bins. Multicalibration strengthens this guarantee by conditioning additionally on a set of groups $\mathcal{G}$, requiring calibration to hold simultaneously within every group $g \in \mathcal{G}$.
Formally, following Hébert-Johnson et al.~\cite{HebertJohnsonK18}, a scoring function $\hat{p}$ is  \emph{$\alpha$-approximately multicalibrated} with respect to a data distribution $\mathcal{D}$ and a set of groups $\mathcal{G}$ iff 

\begin{equation}
\label{eq:multicalibration}
    \text{gASCE}(g) < \frac{\alpha}{\mathbb{P}_\mathcal{D}(g(X)=1)}, \quad \forall g \in \mathcal{G},
\end{equation}

where $\alpha > 0$ is the admitted error.

\section{Research Methodology / Approach}

Our objective is to calibrate  the confidence scores of generative models for code rather than improve their raw accuracy. We focus on function synthesis~\cite{Chen21HumanEval,Austin21MBPP,JainHGLYZWSSS25} as a widely studied, representative task for code generation and aim to develop models that reliably produces confidence scores for the probability that a prompt-completion pair is correct. 

Our methodology involves two steps. First, we evaluate how well different initial uncertainty estimation methods align with the actual correctness of generated outputs. Next, we evaluate and compare several approaches for post hoc calibration of these uncertainty measures. Particularly, we compare \emph{multicalibration} with group-agnostic baselines.

\subsection{Research Agenda}
To systematically assess calibration techniques for code generation across different influencing factors, we define the following research questions:
\begin{itemize}
    \item {\bf RQ1: How effective is multicalibration for post-hoc calibration?} And how does it compare to both uncalibrated confidence scores and group-agnostic baselines? 
    \item {\bf RQ2: How do different initial confidence measures compare?} What impact does the initial scoring method have on the final calibration performance?
    \item {\bf RQ3: Which types of group information contribute most to improved calibration?} 
\end{itemize}

\subsection{Initial Confidence Measures} \label{approach:initial} 
\label{sec:initial_confidence}
A key challenge in our setup is selecting an appropriate scoring method to derive initial confidences $\hat{p}$ as a starting point for calibration. A range of uncertainty quantification techniques for LLMs have been proposed, broadly categorized into black-box and white-box approaches (see \cite{Shorinwa25UQsurvey} for a comprehensive survey). 

However, a recent comparison of confidence measures for generative LLMs~\cite{VashurinFVRVTPXSGPBNPS25} found no clear winner, better-performing methods were often computationally costly with limited gains. Among simpler approaches, confidence-based measures consistently outperformed reflective ones. We therefore adopt this strategy and define our initial confidence score as the average token likelihood, or \emph{inverse perplexity}, of the generation~\cite{jelinek1977perplexity}:
\begin{equation}
    \label{eq:likelihoods}
    \hat{p} = \text{exp} \left ( \frac{1}{n} \sum_{i=1}^n \; \text{log}\; P(z_i \mid z_{<i}, X)\right )
\end{equation}
where $X$ is the prompt and $Z=[z_1, \dots, z_n]$ the generated output.
Note that this output contains not only the generated code, but also additional reasoning traces, since we use current reasoning models for our experiments.
Therefore, we compare aggregations of token-level confidences over different scopes: 
 (1) \emph{avg\_prob}, the mean likelihood over the entire output sequence, as defined in Equation \ref{eq:likelihoods}; (2) \emph{code\_prob}, which is averaged only over the tokens of the generated code; and (3) \emph{tail\_prob}, which focuses on the last $k$ tokens, as reasoning quality is known to decline towards the end of generations, so they indicate correctnesss~\cite{Fu25DeepThink}.

\subsection{Calibration Approaches}
\label{sec:calibration_approaches}
Previous work has demonstrated that LLM-generated confidences contain a signal for output correctness, but they are not well calibrated~\cite{GuoPSW17, JiangADN21,ZhouS024,Spiess0PPRAJDA25}. To address this, we examine various post-hoc calibration strategies that adjust probability estimates to better align with observations. We distinguish between simple rescaling methods, which aim to improve overall calibration (see Section \ref{sec:calibration}) and more informed, group-based methods, which are designed to provide stronger guarantees through multicalibration (Section \ref{sec:multicalibration}).


\subsubsection{Simple Rescaling Approaches}
We evaluate \emph{Platt scaling} and \emph{Histogram Binning} as two common calibration methods.

\paragraph{Platt Scaling~\cite{platt1999probabilistic}} Platt scaling is a common approach that applies logistic regression to the prediction logits, $log \; \hat{p}$.  
More specifically, in the context of neural networks~\cite{NiculescuMizilC05}, Platt scaling learns scalar parameters $a, b \in \mathbb{R}$ from a training set of confidence-correctness pairs $(\hat{p}_i,y_i)$. It outputs calibrated probabilities by applying the sigmoid function $\sigma$ to the scaled log-probabilities, obtaining supposedly calibrated confidences
\begin{equation}
\label{eq:platt}
\hat{p}^{PLATT} := \sigma(a \cdot \text{log}\;\hat{p} + b).
\end{equation}
Since Spiess et al.~\cite{Spiess0PPRAJDA25} found Platt scaling to improve calibration for code LLMs with some caveats, we include it as a baseline.

\paragraph{Histogram Binning~\cite{ZadroznyE01}} Histogramm Binning is a simple non-parametric calibration method. Following Section \ref{sec:calibration}, it first partitions $[0,1]$ into uniform bins $B_1, \dots, B_M$ 
and 
rounds the raw confidences $\hat{p}$ to the closest grid value $\hat{p}'$:

\begin{equation}
\label{eq:binning}
    \hat{p}' := \underset{b \in [\frac{1}{m}]}{\text{arg min}} \; |\hat{p} - b|. 
\end{equation}

Then, it computes the calibration error $ \Delta_m$ per bin as
\begin{equation}
    \Delta_m := \frac{1}{|B_m|} \cdot \sum_{i\in B_m}y_i-\hat{p}'_i
\end{equation}
and applies it as a constant shift to all samples in the respective bin:
\begin{equation}
\label{eq:hb}
    \hat{p}^{HB} := \hat{p}_i' + \Delta_m \quad \forall m \in (1, \dots M) \quad \forall i \in B_m  \\
\end{equation}

\subsubsection{Multicalibration-based Approaches}
As described in Section \ref{sec:multicalibration}, multicalibration aims at
taking  particular characteristics of the prompt / generation pairs into account, by assigning them to semantic {\it groups}. We first specify the group functions $g$ we use to capture properties of coding problems. We then describe the group-aware multicalibration approaches we test (namely, Group-Conditional Unbiased Regression (GCUR), Iterative Grouped Histogram Binning (IGHB) and Iterative Grouped Linear Binning (IGLB)).  

\paragraph{Group Functions}
Multicalibration is defined with respect to a set of \emph{group functions}. 
If these groups capture features correlated with 
the likelihood of correctness, their inclusion is likely to improve calibration. At the same time, to ensure practical applicability, we restrict ourselves to features that can be computed efficiently.

Specifically, we define the following group functions: (1) \emph{complex-ity-based groups} aim to capture differences in code complexity by grouping examples based on circular complexity or, where available, on benchmark-provided difficulty labels. (2) \emph{length-based groups} divide examples according to the output length, measured either in total characters or lines of code of the extracted code snippet. (3) \emph{language-based groups} group examples by the programming language of the generated code. Based on these groups, we evaluate the following methods:

\paragraph{Group-Conditional Unbiased Regression (GCUR)~\cite{GopalanHKRW23}} GCUR provides a \emph{group conditionally unbiased} model as defined in Equation~\ref{eq:multiaccuracy}. It uses Algorithm~\ref{alg:GCUR} to produce a score $\hat{p}^{GCUR}$ 
by solving a linear regression problem over features defined by the raw score $\hat{p}$ and the group indicator functions in $\mathcal{G}$.

\begin{algorithm}
    \caption{Group-Conditional Unbiased Regression}
    \label{alg:GCUR}
\begin{algorithmic}[1]

\State Set \hspace{0.5em} \parbox[t]{.7\linewidth}{
$\hat{p}^{GCUR} := \hat{p} + \sum_{g \in \mathcal{G}} \lambda_g  \cdot g(x),$\\
\vspace{0.5em} \\
s.t. $\{\lambda_g\}_{g \in \mathcal{G}} = \text{arg min }\mathcal{B}(\hat{p}^{GCUR})$}

\end{algorithmic}
\end{algorithm}
Gopalan et al.~\cite{GopalanHKRW23} later showed that the condition also holds if MSE (or Brier Score $\mathcal{B}$) is replaced with a cross-entropy loss, i.e. when solving a logistic regression instead of a linear regression. We evaluate both variants, denoted as $\hat{p}^{LINR}$ and $\hat{p}^{LOGR}$ respectively.

\paragraph{Iterative Grouped Histogram Binning (IGHB)~\cite{HebertJohnsonK18, DetommasoLF024}} IGHB (Algorithm \ref{alg:ighb}) extends Histogram Binning by the concept of groups. The method first evaluates whether $\alpha$-approximate multicalibration (as defined in Equation \ref{eq:multicalibration}) is satisfied. 
If this is not the case, IGHB identifies the group/bin combination that contributes most to the gASCE (Step 3) and patches the confidences of the training samples within this bin/group combination (Step 4). IGHB then iterates until convergence. The rounding operation $h'$ in Step 5 (as defined in Equation~\ref{eq:binning}) ensures that the number of bins does not increase, guaranteeing sufficient data in each step. 
Note that as the confidences $\hat{p}_i$ are corrected over the iterations of the algorithm, the memberships to bins $B_m$ (and thus $P(g,m),\Delta_{g,m}$, and $gASCE(g)$) change. This is why we include an additional  parameter $t$.

\begin{algorithm}
    \caption{Iterative Grouped Histogram Binning (IGHB)}
    \label{alg:ighb}
\begin{algorithmic}[1]
\State Let $\alpha = \frac{1}{M}, t=0, \; \hat{p}_{i}^{t=0} := \hat{p}_i'$ for all $i=1,...,n$
\vspace{0.5em}
\While{$\text{max}_{g \in\mathcal{G}} \; P(g) \cdot gASCE^t(g) > \alpha$}
\vspace{0.5em}
\State Set $(g^*,m^*) := \underset{g \in \mathcal{G}, m \in \{1,...,M\}}{\text{arg max}} \; P^t(g,m) \cdot (\Delta_{g,m}^t)^2$
\vspace{0.5em}
\State Set \parbox[t]{.8\linewidth}{
      \hspace{0pt}%
      $h_{t+1} := \left\{\begin{array}{l}
                                \hat{p}_i^t + \Delta_{g^*,m^*}^t \quad \text{if } i \in  B_{m^*}^t {\cap} G^*, \\
                                \hat{p}_i^t \hspace{5.2 em}  \text{otherwise.}
                            \end{array}\right.$}
\vspace{0.5em}
\State Set $\hat{p}^{t+1} := (h^{t+1})'$ 
\EndWhile

\end{algorithmic}
\end{algorithm}

\paragraph{Iterative Grouped Linear Binning (IGLB)~\cite{DetommasoLF024}} IGHB, as introduced above, can build complex models by performing iterative updates on intersecting groups, which renders the method prone to overfitting~\cite{GlobusHarrisHK23}. One reason is that IGHB operates on group-bin combinations that may contain only a small number of data points, leading to inaccurate estimates of the distribution. To mitigate this issue, Detommaso et al.~\cite{DetommasoLF024} propose \emph{Iterative Grouped Linear Binning}, demonstrating that -- instead of splitting the data into disjoint bins as defined in Equation~\ref{eq:bins} -- it is beneficial to use larger, overlapping bins $B_m^\le = \{ i \mid \hat{p}_i \le \frac{m}{M}\}$ (and  $B_m^\ge$ analogously) to obtain more stable estimates. Since these bins now contain a range of values rather than sets of identical values, IGLB extends the update step (Step 4 in Algorithm~\ref{alg:ighb}) to \emph{linear patches} of the form $h_{t+1} = \alpha + \beta \cdot \hat{p}^t$ where $\alpha$ and $\beta$ are chosen so as to minimize the squared error of the model. Specifically, linear scaling (LS) is applied as: 
\begin{align}
\label{eq:ls}
    \text{LS}(\hat{p}) :&= \sigma(\alpha^* + \beta^*\, \text{logit}(\hat{p})),\\
    \text{with} \quad(\alpha^*, \beta^*) &= \underset{\alpha, \beta}{\text{arg min}} \;\mathcal{B}(\hat{p}) \nonumber
\end{align}
Algorithm~\ref{alg:iglb} integrates these strategies into IGHB by iteratively selecting the group-bin combination with the highest error (Step 3) and applying the LS patch according to Equation~\ref{eq:ls} (Step 5). To prevent overfitting, the  method includes two early-stopping rules: it halts if the Brier Score $\mathcal{B}_{val}$ on a held out validation set no longer improves (Step 6) or if the probability mass of the conditioning event selected for an update is less than a predefined threshold.

\begin{algorithm}
    \caption{Iterative Grouped Linear Binning (IGLB)}
    \label{alg:iglb}
\begin{algorithmic}[1]
\State Let $t=0, \epsilon>0$, \; $\hat{p}_{i}^{t=0} := \hat{p}_i'$ for all $i=1,...,n$
\vspace{0.5em}
\While{True}
\vspace{0.5em}
\State Set $(g^*, m^*, \tau^*) := \underset{\substack{g \in \mathcal{G},\\ m \in \{1,\dots,M\},\\  \tau \in \{\le, \ge\}}}{\text{arg max}}  P^t(g,m, \tau) \cdot (\Delta_{g,m,\tau}^{t})^2$
\vspace{0.5em}
\State{\textbf{break} if $ P^t(g^*,m^*, \tau^*) < \epsilon$}
\vspace{0.5em}
\State Set \parbox[t]{.8\linewidth}{
      \hspace{0pt}%
      $h_{t+1} := \left\{\begin{array}{l}
                                \text{LS}(\hat{p}_i^t) \quad \text{if } i \in B_{m^*}^{\tau^*} {\cap} G^*, \\
                                \hat{p}_i^t \hspace{3 em}  \text{otherwise.}
                            \end{array}\right.$}
\vspace{0.5em}
\State{\textbf{break} if $\mathcal{B}_{\mathcal{D}_{val}}(h_{t+1},Y) \ge \mathcal{B}_{\mathcal{D}_{val}}(\hat{p}^t,Y)$}
\vspace{0.5em}
\State Set $\hat{p}_{t+1} := h'_{t+1}$
\vspace{0.5em}
\EndWhile

\end{algorithmic}
\end{algorithm}

\begin{table*}[ht!]
\caption{Comparison of Brier Skill Score (BSS) and binary clasification accuracy (ACC) across uncalibrated scores, two group-agnostic baselines, and four group-based calibration methods. Results for DeepSeek R1 distill on LiveCodeBench are excluded due to unreliable code extraction.}
\label{tab:results_comparison}
\begin{tabular}{llrrrrrrrrrrrr}
\toprule
&   &      & \multicolumn{3}{c}{\textbf{LiveCodeBench}}                                           && \multicolumn{3}{c}{\textbf{MultiPL-E}}                                                                & & \multicolumn{3}{c}{\textbf{McEval}}                                                                                                                           \\ \cmidrule(rl){4-6}\cmidrule(rl){8-10}\cmidrule(l){12-14}
\textbf{BSS} & method & groups     &\textit{ Qwen3  }                    & \textit{gpt-oss}                            & \textit{R1(dist)} &&\textit{ Qwen3  }                            & \textit{gpt-oss }                          & \textit{R1(dist)      }                 && \textit{Qwen3  }                            &\textit{ gpt-oss   }                        & \textit{R1(dist)}                       \\ \midrule
&uncalibrated & - & \LCBQwenIIIavgprobUncalibSkillScore & \LCBGPTOSSavgprobUncalibSkillScore &     -     && \HEQwenIIIavgprobUncalibSkillScore & \HEGPTOSSavgprobUncalibSkillScore & \HEDSRavgprobUncalibSkillScore && \MEQwenIIIavgprobUncalibSkillScore & \MEGPTOSSavgprobUncalibSkillScore & \MEDSRavgprobUncalibSkillScore \\
&Platt    &  - & \LCBQwenIIIavgprobPLATTSkillScore & \LCBGPTOSSavgprobPLATTSkillScore &     -     && \HEQwenIIIavgprobPLATTSkillScore & \HEGPTOSSavgprobPLATTSkillScore & \HEDSRavgprobPLATTSkillScore && \MEQwenIIIavgprobPLATTSkillScore & \MEGPTOSSavgprobPLATTSkillScore & \MEDSRavgprobPLATTSkillScore \\ 
&HB       & - & \LCBQwenIIIavgprobHBSkillScore      & \LCBGPTOSSavgprobHBSkillScore      &    -      && \HEQwenIIIavgprobHBSkillScore      & \HEGPTOSSavgprobHBSkillScore      & \HEDSRavgprobHBSkillScore      && \MEQwenIIIavgprobHBSkillScore      & \MEGPTOSSavgprobHBSkillScore      & \MEDSRavgprobHBSkillScore      \\
&LINR     & \checkmark & \LCBQwenIIIavgprobLRSkillScore      & \LCBGPTOSSavgprobLRSkillScore      &     -     && \textbf{\HEQwenIIIavgprobLRSkillScore} & \textbf{\HEGPTOSSavgprobLRSkillScore}& \textbf{\HEDSRavgprobLRSkillScore}&& \MEQwenIIIavgprobLRSkillScore& \textbf{\MEGPTOSSavgprobLRSkillScore}& \MEDSRavgprobLRSkillScore      \\
&LOGR     & \checkmark & \LCBQwenIIIavgprobLOGRSkillScore    & \LCBGPTOSSavgprobLOGRSkillScore    &    -      && \HEQwenIIIavgprobLOGRSkillScore    & \HEGPTOSSavgprobLOGRSkillScore    & \HEDSRavgprobLOGRSkillScore    && \MEQwenIIIavgprobLOGRSkillScore    & \MEGPTOSSavgprobLOGRSkillScore    & \MEDSRavgprobLOGRSkillScore    \\
&IGHB     & \checkmark & \LCBQwenIIIavgprobIGHBSkillScore    & \LCBGPTOSSavgprobIGHBSkillScore    &   -       && \HEQwenIIIavgprobIGHBSkillScore    & \HEGPTOSSavgprobIGHBSkillScore    & \HEDSRavgprobIGHBSkillScore    && \MEQwenIIIavgprobIGHBSkillScore    & \MEGPTOSSavgprobIGHBSkillScore    & \MEDSRavgprobIGHBSkillScore    \\
&IGLB     & \checkmark & \textbf{\LCBQwenIIIavgprobIGLBSkillScore}&\textbf{\LCBGPTOSSavgprobIGLBSkillScore}& - && \HEQwenIIIavgprobIGLBSkillScore    & \HEGPTOSSavgprobIGLBSkillScore    & \HEDSRavgprobIGLBSkillScore    && \textbf{\MEQwenIIIavgprobIGLBSkillScore}    & \MEGPTOSSavgprobIGLBSkillScore   & \textbf{\MEDSRavgprobIGLBSkillScore}     \\ \cmidrule(rl){4-6}\cmidrule(rl){8-10}\cmidrule(l){12-14}
\textbf{ACC}&  &     & \textit{Qwen3 }                     & \textit{gpt-oss }                           &\textit{ R1(dist)}  &&\textit{ Qwen3  }                            &\textit{ gpt-oss }                          & \textit{R1(dist) }                      && \textit{Qwen3  }                                           & \textit{gpt-oss}                                          & \textit{R1(dist) }                                     \\ \midrule
&uncalibrated & - & \LCBQwenIIIavgprobUncalibACC        & \LCBGPTOSSavgprobUncalibACC        &    -      && \HEQwenIIIavgprobUncalibACC        & \HEGPTOSSavgprobUncalibACC        & \HEDSRavgprobUncalibACC        && \MEQwenIIIavgprobUncalibACC        & \MEGPTOSSavgprobUncalibACC        & \MEDSRavgprobUncalibACC        \\  
&Platt    & - & \LCBQwenIIIavgprobPLATTACC          & \LCBGPTOSSavgprobPLATTACC        &    -      && \HEQwenIIIavgprobPLATTACC        & \HEGPTOSSavgprobPLATTACC        & \HEDSRavgprobPLATTACC        && \MEQwenIIIavgprobPLATTACC        & \MEGPTOSSavgprobPLATTACC        & \MEDSRavgprobPLATTACC        \\ 
&HB       & - & \LCBQwenIIIavgprobHBACC             & \LCBGPTOSSavgprobHBACC             &    -      && \HEQwenIIIavgprobHBACC             & \HEGPTOSSavgprobHBACC             & \HEDSRavgprobHBACC             && \MEQwenIIIavgprobHBACC             & \MEGPTOSSavgprobHBACC             & \MEDSRavgprobHBACC             \\
&LINR     & \checkmark & \LCBQwenIIIavgprobLRACC             & \LCBGPTOSSavgprobLRACC             &   -       && \textbf{\HEQwenIIIavgprobLRACC}    & \HEGPTOSSavgprobLRACC             & \HEDSRavgprobLRACC             && \MEQwenIIIavgprobLRACC             & \textbf{\MEGPTOSSavgprobLRACC}    & \textbf{\MEDSRavgprobLRACC}    \\
&LOGR     & \checkmark & \LCBQwenIIIavgprobLOGRACC           & \textbf{\LCBGPTOSSavgprobLOGRACC}  &   -       && \HEQwenIIIavgprobLOGRACC           & \textbf{\HEGPTOSSavgprobLOGRACC}  & \textbf{\HEDSRavgprobLOGRACC}  && \textbf{\MEQwenIIIavgprobLOGRACC}  & \MEGPTOSSavgprobLOGRACC           & \MEDSRavgprobLOGRACC           \\
&IGHB     & \checkmark & \LCBQwenIIIavgprobIGHBACC           & \LCBGPTOSSavgprobIGHBACC           &    -      && \HEQwenIIIavgprobIGHBACC           & \HEGPTOSSavgprobIGHBACC           & \HEDSRavgprobIGHBACC           && \MEQwenIIIavgprobIGHBACC           & \MEGPTOSSavgprobIGHBACC           & \MEDSRavgprobIGHBACC           \\
&IGLB     & \checkmark & \textbf{\LCBQwenIIIavgprobIGLBACC}  & \LCBGPTOSSavgprobIGLBACC           &    -      && \HEQwenIIIavgprobIGLBACC           & \HEGPTOSSavgprobIGLBACC           & \HEDSRavgprobIGLBACC           && \MEQwenIIIavgprobIGLBACC           & \MEGPTOSSavgprobIGLBACC           & \MEDSRavgprobIGLBACC           \\ \bottomrule
\end{tabular}
\end{table*}

\section{Experimental Setup}

This section outlines the task setup, datasets and models used in our study.
We conduct calibration experiments on \emph{function synthesis}, which aims to generate executable methods from natural language problem descriptions (see Subsection \ref{sec:benchmarks} for details on the benchmarks). We prompt three recent open-weight reasoning models (see Subsection~\ref{sec:models}) to produce ten generations per sample, recording the associated token probabilities. This yields $171{,}420$ prompt–generation–probability tuples. Generated code is evaluated via benchmark unit tests and labeled as correct if all tests pass.

Using this dataset, we compare three initial confidence measures as described in Section~\ref{sec:initial_confidence} and six calibration methods: two classical, two group-accuracy-based, and two multicalibration methods (Section~\ref{sec:calibration_approaches}). All experiments use a grid of $M=20$ bins, which preliminary tests found to balance granularity and bin weights effectively.

\subsection{Benchmarks}
\label{sec:benchmarks}
To capture varying levels of difficulty and language coverage, we employ three complementary benchmarks: MultiPL-E~\cite{CassanoGNNPPYZAFGGJ23}, McEval~\cite{ChaiL0YJLS0RGWW25} and LiveCodeBench~\cite{JainHGLYZWSSS25}.

\textit{MultiPL-E~\cite{CassanoGNNPPYZAFGGJ23}} is a multilingual code generation benchmark created by translating two popular Python benchmarks (HumanEval~\cite{Chen21HumanEval} and MBPP~\cite{Austin21MBPP}) into 18 programming languages. For our study, we use its 2,952 samples based on HumanEval.

\textit{McEval~\cite{ChaiL0YJLS0RGWW25}} is a multilingual code benchmark comprising 16K test samples across 40 languages, covering varying difficulty levels and coding scenarios. We use its multilingual code-generation subset of 1,707 samples, which was constructed from curated code snippets augmented with LLM-generated instructions and test cases.

\textit{LiveCodeBench~\cite{JainHGLYZWSSS25}} is a collection of recently published Python coding problems from three competition platform, designed to provide \emph{contamination free evaluation} of LLMs for code, i.e. addressing the issue of data leakage, where parts of the evaluation data may be included in a model's training corpus. We use its most recent release (release\_v6) containing 1055 problems from May 2023 to April 2025. 

The three benchmarks complement each other in scope and difficulty. MultiPL-E and McEval provide multilingual problems -- MultiPL-E focusing on relatively simple tasks, and McEval covering a broader range of difficulty levels. LiveCodeBench, in contrast, features more challenging and diverse Python problems, and enables contamination-free evaluation, thereby mitigating potential calibration biases caused by training data overlap. 

All datasets include executable test suites for functional verification, which we adopt as the most reliable measure of code correctness. 
For all experiments, we require a training set for calibration, a hold-out test set, and an additional validation set for IGLB calibration. To prevent leakage across splits, we partition the data at the \emph{problem level}, ensuring that all generations and translations of a given problem reside within the same split.

\subsection{Models}
\label{sec:models}
We evaluate three recent open-weight reasoning LLMs in our experiments: \emph{Qwen3-Coder-30B-A3B-Instruct}~\cite{yang25qwen3}, \emph{OpenAI GPT-OSS-20B}~\cite{openai2025gptoss120bgptoss20bmodel} and \emph{DeepSeek-R1-Distill-Qwen-32B}~\cite{guo25deepseek}. Qwen3-Coder is a family of Mixture-of-Experts (MoE) models for coding and agentic tasks, we use the 30B-A3B-Instruct variant. The GPT-OSS models from OpenAI are MoE models for general reasoning and agentic tasks, and we employ the 20b variant in our evaluation. DeepSeek-R1-Distill-Qwen-32B is a Qwen-32B model distilled from DeepSeek-R1, a large reasoning model.  
All three models demonstrate strong coding performance, and are open-weight, which is essential for our use-case as we need access to token-level generation probabilities.

\begin{figure*}[ht]
    \centering
    \includegraphics[width=1\linewidth]{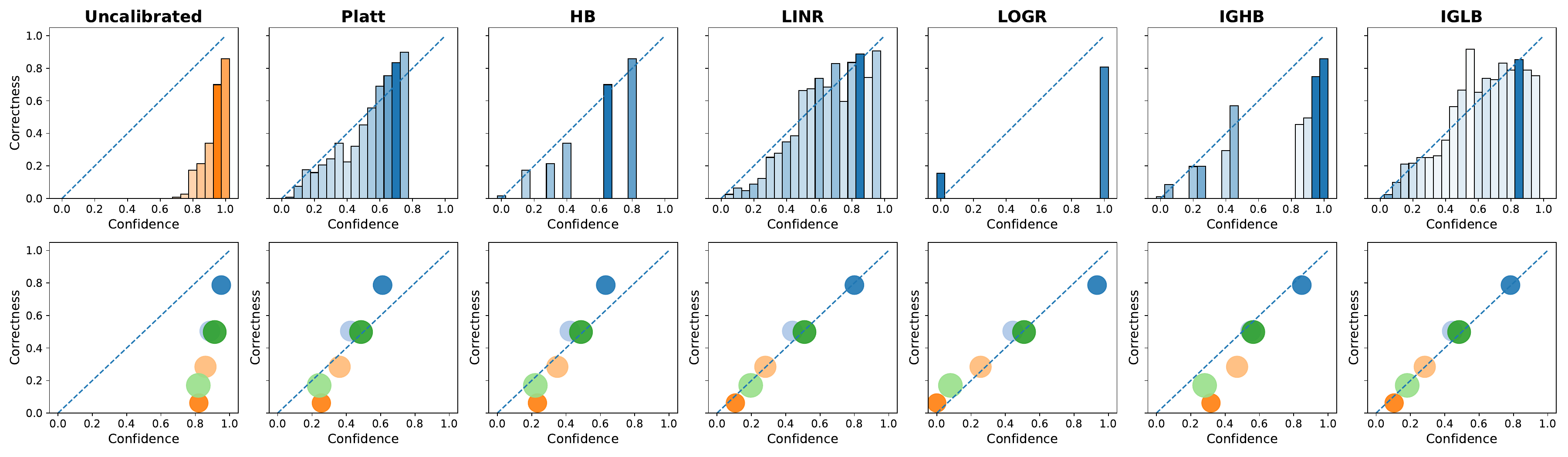}
    \caption{\emph{Top}: Reliability diagrams for Qwen3 Coder on LiveCodeBench. Well calibrated models fit the identity line, while deviations from the diagonal indicate miscalibration. Bin size is encoded by color intensity. \emph{Bottom}: Mean predicted scores versus accuracies across clustered groups. LINR and IGLB achieve the best calibration, consistent with Table~\ref{tab:results_comparison}.}
    \label{fig:bar_charts}
\end{figure*}

\section{Results}

We first provide an overall quantitative impression of all calibration methods, datasets  and models (Section \ref{results:overview}), then discuss the influence of the different initial scoring methods (Section \ref{results:initial}), followed by a more detailed look on groups, with an ablation study and group-specific calibration results (Section \ref{results:groups}).

\subsection{RQ1: How effective is multicalibration for post-hoc calibration?} \label{results:overview}
We compare the uncalibrated initial scoring function (using \emph{avg\_score}) against its calibrated variants obtained using the above post-processing algorithms: Platt scaling (Equation~\ref{eq:platt}), Histogram Binning (HB, Equation~\ref{eq:hb}), the linear and logistic regression variants of Group-Conditional Unbiased Regression (Algorithm~\ref{alg:GCUR}, LINR and LOGR), Iterative Grouped Histogram Binning (Algorithm~\ref{alg:ighb}, IGHB) and Iterative Grouped Linear Binning (Algorithm~\ref{alg:iglb}, IGLB). While Platt scaling and HB perform standard calibration without requiring a grouping strategy. LINR and LOGR ensure group-conditional unbiasedness (Equation~\ref{eq:multiaccuracy}), and IGHB and IGLB produce multicalibrated models as defined in Equation~\ref{eq:multicalibration}.

Table~\ref{tab:results_comparison} reports the Brier Skill Score (BSS) and binary classification accuracy (ACC) -- thresholding at $\hat{p}{\geq}50\%$ -- across all methods, benchmarks and LLMs.
Note that results for DeepSeek R1 distill on LiveCodeBench are omitted, due to formatting inconsistencies that prevented reliable code extraction.

Across all datasets and models, IGLB and the two Group-Conditio-nal Unbiased Regression variants (LINR and LOGR) consistently achieve the best performance. All calibration methods improve upon the uncalibrated baseline, in many cases substantially. While uncalibrated scores were often near-random ($\text{ACC} \approx 0.5$) for distinguishing correct from incorrect code, all post-hoc methods yielded substantial gains. After calibration, accuracy rose to between $72.5\%$ and $95.5\%$ across all models and datasets -- an improvement of up to $58.4\%$ in the most notable case (DeepSeek R1 distill on McEval). This highlights the potential of calibrated uncertainty measures to effectively predict the correctness of LLM-generated code even without unit tests.

Standard calibration methods such as Platt-Scaling and HB perform well compared to the initial scoring function, yet consistently underperform the multicalibrated IGLB method and the multiaccurate LINR/LOGR. Figure~\ref{fig:bar_charts} visualizes sample results as reliability diagrams, plotting confidence scores against empirical accuracy (top). 
The bottom panels plot the averaged confidence scores against the fraction of positive labels across various groups (each color representing one group).
All calibration methods improve upon the overconfident, uncalibrated scores (left) with IGLB and LINR demonstrating the strongest calibration performance.

\subsection{RQ2: How do different initial confidence measures compare?} \label{results:initial}

\begin{table}[b]
\caption{Comparison of initial scoring methods for Qwen3 Coder on LiveCodeBench, reporting the Expected Calibration Error (ECE, lower is better) and Brier Skill Score ($BSS$, higher is better). 
For fair comparison, examples where no code could be extracted from the generated output are excluded.}
\label{tab:probs}
\begin{tabular}{lrrrrrr}
\toprule
      & \multicolumn{2}{r}{\textbf{avg\_prob}}                     & \multicolumn{2}{r}{\textbf{code\_prob}}                     & \multicolumn{2}{r}{\textbf{tail\_prob}}                      \\ \cmidrule(r){2-3} \cmidrule(lr){4-5} \cmidrule(l){6-7}
\textbf{Qwen3 C.}        & ECE           & $BSS$                           & ECE                           & $BSS$ & ECE                            & $BSS$ \\ \midrule
uncalib. & \bf \LCBfQwenIIIavgprobUncalibECE & \bf \LCBfQwenIIIavgprobUncalibSkillScore & \LCBQwenIIIcodeprobUncalibECE        & \LCBQwenIIIcodeprobUncalibSkillScore & \LCBfQwenIIItailprobUncalibECE & \LCBfQwenIIItailprobUncalibSkillScore \\
HB       & \LCBfQwenIIIavgprobHBECE     & \bf \LCBfQwenIIIavgprobHBSkillScore      & \bf \LCBQwenIIIcodeprobHBECE      & \LCBQwenIIIcodeprobHBSkillScore     & \LCBfQwenIIItailprobHBECE      & \LCBfQwenIIItailprobHBSkillScore      \\
LINR     &  \bf \LCBfQwenIIIavgprobLRECE & \bf \LCBfQwenIIIavgprobLRSkillScore  & \LCBQwenIIIcodeprobLRECE    & \LCBQwenIIIcodeprobLRSkillScore   & \LCBfQwenIIItailprobLRECE    & \LCBfQwenIIItailprobLRSkillScore    \\
LOGR     &  \bf \LCBfQwenIIIavgprobLOGRECE&\bf \LCBfQwenIIIavgprobLOGRSkillScore& \LCBQwenIIIcodeprobLOGRECE    & \LCBQwenIIIcodeprobLOGRSkillScore   & \LCBfQwenIIItailprobLOGRECE    & \LCBfQwenIIItailprobLOGRSkillScore    \\
IGLB     & \bf \LCBfQwenIIIavgprobIGLBECE& \bf \LCBfQwenIIIavgprobIGLBSkillScore& \LCBQwenIIIcodeprobIGLBECE    & \LCBQwenIIIcodeprobIGLBSkillScore   & \LCBfQwenIIItailprobIGLBECE    & \LCBfQwenIIItailprobIGLBSkillScore    \\ \bottomrule
\end{tabular}
\end{table}

Table~\ref{tab:probs} compares calibration results for the three initial confidence-based scoring methods derived from different aggregation strategies (Section \ref{approach:initial}). \emph{avg\_prob} computes the mean likelihood over the entire output sequence, including possible reasoning traces, \emph{code\_prob} averages only over the tokens of the generated code, and \emph{tail\_prob} averages over the last $40$ tokens of the generation.
To ensure a fair comparison, we exclude examples where no code could be extracted, as code is required for \emph{code\_prob}. 

We report results only for Qwen3 Coder, since GPT-OSS primarily failed in cases where no executable code was produced. Since -- as mentioned above -- we filtered these cases in this experiment, GPT-OSS produced mostly correct generations ($\LCBfGPTOSSavgprobUncalibACC$) compared to $\LCBfQwenIIIavgprobUncalibACC$ for Qwen3 Coder. As GPT-OSS either produced correct or no code, we consider this experiment uninformative for that model.

For Qwen3 Coder, \emph{avg\_prob} achieves the best calibration results, both for the uncalibrated scores and after post-hoc calibration. The uncalibrated ECE for \emph{avg\_prob} was $0.048$ and $0.052$ lower than for \emph{code\_prob} and \emph{tail\_prob}, respectively. For all group-based calibration methods, we achieve the best ECE and BSS using \emph{avg\_prob} as the initial score. For instance, after IGLB calibration, the ECE remained $0.024$ lower than with the other scores. 
This supports the intuition that the entire reasoning trace provides a valuable confidence signal, while the hypothesis of the generation tail being most relevant was not confirmed. 
Based on these findings, we argue for \emph{avg\_prob} as a robust initial score that can be easily obtained from any LLM that provides access to its token likelihoods, and adopt it as the baseline in all further experiments. 

\subsection{RQ3: Which types of group information contribute most to improved calibration?} \label{results:groups}
Table~\ref{tab:gasce_per_group} reports the group average squared error (gASCE) per group for the best performing methods from Table~\ref{tab:results_comparison}. While the group-conditional unbiased regression methods (LINR and LOGR) often achieved the highest accuracy in Table~\ref{tab:results_comparison}, IGLB yields the lowest gASCE for most groups. This observation aligns with theoretical expectations: LINR and LOGR ensure low group-conditional bias but do not guarantee calibration \emph{within} each group, which is precisely what the gASCE metric captures. The bottom plots in Figure~\ref{fig:bar_charts} illustrate this observation: average confidence scores per group plotted against their  fraction of correct predictions align most closely with the diagonal for IGLB, while LINR and LOGR show slight deviations from the diagonal. 

\begin{figure}
    \centering
    \includegraphics[width=1\linewidth]{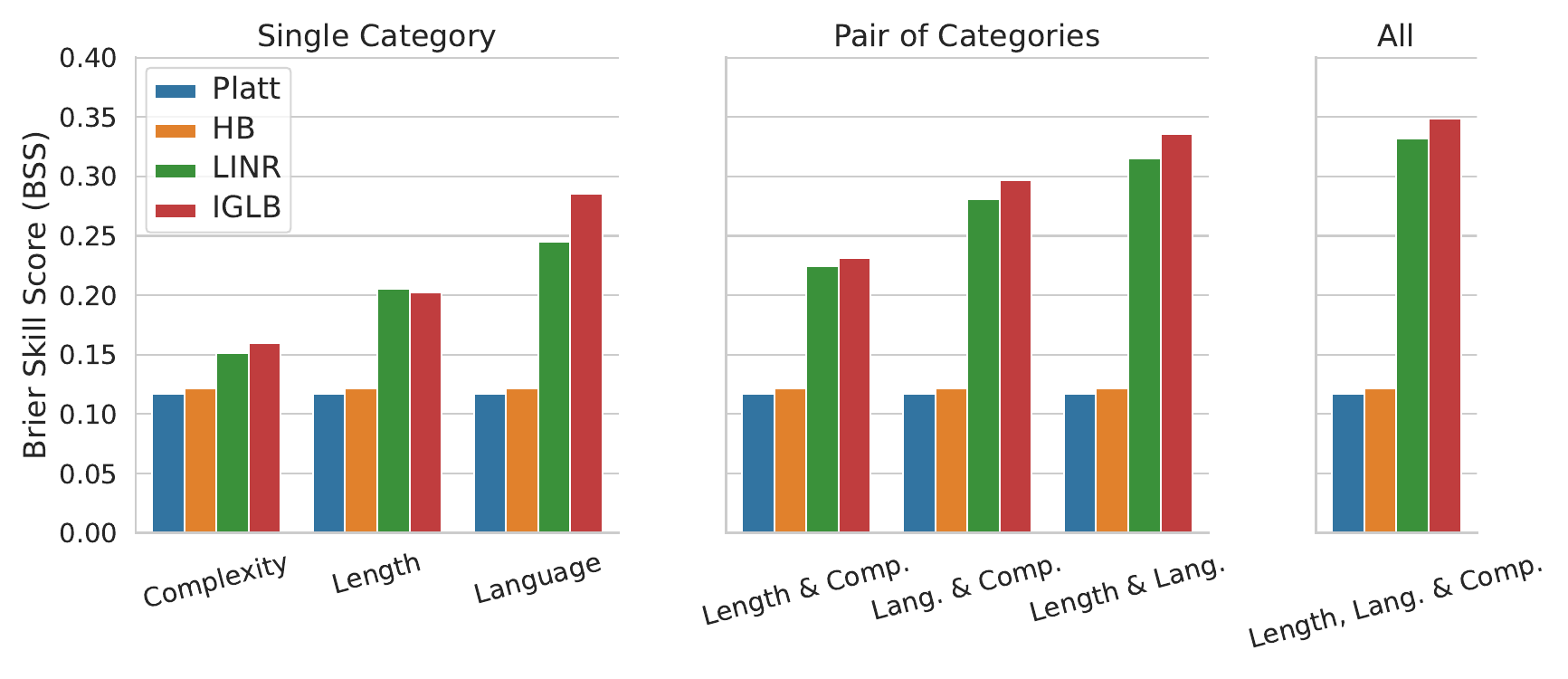}
    \caption{Ablation experiments evaluating different group combinations on McEval. Results are averaged over the three LLMs.}
    \label{fig:ablation}
\end{figure}

\paragraph{Ablation experiments} To assess the impact of different grouping features from Section~\ref{sec:groups}, we performed ablations on McEval, omitting individual group categories and evaluating their effect on calibration. Figure~\ref{fig:ablation} reports the mean $BSS$ across all models, for the best-performing methods from Table~\ref{tab:results_comparison} including group-agnostic methods for comparison. 
When using only a single category (\emph{complexity}, \emph{language} or \emph{length}), \emph{language} provides the largest performance gain while \emph{complexity} contributes the least. Interestingly, among the combinations of two categories, \emph{complexity \& length} yield the lowest results, suggesting that these categories contain redundant information, i.e. that in some cases, code length implicitly encodes information about code complexity and vice versa. The best-performing pair is \emph{language \& length}, meaning that omitting the \emph{complexity} related groups is least harmful in our experiments. Overall, the highest performance is achieved when all three categories are included, indicating that, despite the previously mentioned redundancies, the categories still provide complementary information.


\begin{table}[h]

\caption{Comparison of Group averaged Squared Error (gASCE) for complexity- and length-based groups and some language-based ones for Qwen3 Coder on McEval. }
\label{tab:gasce_per_group}
\begin{tabularx}{\columnwidth}{Xlrrrrr@{}}
\toprule
gASCE              &type   & uncalib.                               & Platt                               & LINR                             & LOGR                                                            & IGLB                               \\ \midrule
\textit{low}    &comp.& \MEQwenIIIavgprobUncalibgASCEcompeasy  &\MEQwenIIIavgprobPLATTgASCEcompeasy  & \textbf{\MEQwenIIIavgprobLRgASCEcompeasy} & \MEQwenIIIavgprobLOGRgASCEcompeasy &  \textbf{\MEQwenIIIavgprobIGLBgASCEcompeasy} \\
\textit{medium}&comp. & \MEQwenIIIavgprobUncalibgASCEcompmedium&\MEQwenIIIavgprobPLATTgASCEcompmedium& \MEQwenIIIavgprobLRgASCEcompmedium & \MEQwenIIIavgprobLOGRgASCEcompmedium  & \textbf{\MEQwenIIIavgprobIGLBgASCEcompmedium}      \\
\textit{high}  &comp.  & \MEQwenIIIavgprobUncalibgASCEcomphard &\MEQwenIIIavgprobPLATTgASCEcomphard  & \MEQwenIIIavgprobLRgASCEcomphard & \MEQwenIIIavgprobLOGRgASCEcomphard & \textbf{\MEQwenIIIavgprobIGLBgASCEcomphard}      \\
\textit{len\_low}    &length & \MEQwenIIIavgprobUncalibgASCElenlow    &\textbf{\MEQwenIIIavgprobPLATTgASCElenlow} & \textbf{\MEQwenIIIavgprobLRgASCElenlow} & \MEQwenIIIavgprobLOGRgASCElenlow & \textbf{\MEQwenIIIavgprobIGLBgASCElenlow}      \\
\textit{len\_high}   &length & \MEQwenIIIavgprobUncalibgASCElenhigh   &\MEQwenIIIavgprobPLATTgASCElenhigh   & \MEQwenIIIavgprobLRgASCElenhigh & \MEQwenIIIavgprobLOGRgASCElenhigh & \textbf{\MEQwenIIIavgprobIGLBgASCElenhigh}      \\
\textit{loc\_low}    &length & \MEQwenIIIavgprobUncalibgASCEloclow    &\MEQwenIIIavgprobPLATTgASCEloclow    & \textbf{\MEQwenIIIavgprobLRgASCEloclow} & \MEQwenIIIavgprobLOGRgASCEloclow &  \MEQwenIIIavgprobIGLBgASCEloclow      \\                
\textit{loc\_high}   &length & \MEQwenIIIavgprobUncalibgASCElochigh   &\MEQwenIIIavgprobPLATTgASCElochigh & \MEQwenIIIavgprobLRgASCElochigh & \MEQwenIIIavgprobLOGRgASCElochigh & \textbf{\MEQwenIIIavgprobIGLBgASCElochigh}      \\ 
\textit{c}     &lang. & \MEQwenIIIavgprobUncalibgASCElangC     &\MEQwenIIIavgprobPLATTgASCElangC     & \MEQwenIIIavgprobLRgASCElangC & \bf \MEQwenIIIavgprobLOGRgASCElangC & \MEQwenIIIavgprobIGLBgASCElangC      \\ 
\textit{python}&lang. & \MEQwenIIIavgprobUncalibgASCElangPython&\MEQwenIIIavgprobPLATTgASCElangPython& \MEQwenIIIavgprobLRgASCElangPython & \MEQwenIIIavgprobLOGRgASCElangPython & \bf \MEQwenIIIavgprobIGLBgASCElangPython      \\ 
\textit{lua}   &lang. & \MEQwenIIIavgprobUncalibgASCElangLua   &\MEQwenIIIavgprobPLATTgASCElangLua   & \MEQwenIIIavgprobLRgASCElangLua & \textbf{\MEQwenIIIavgprobLOGRgASCElangLua} & \MEQwenIIIavgprobIGLBgASCElangLua      \\ 
\textit{rust}  &lang. & \MEQwenIIIavgprobUncalibgASCElangRust  &\MEQwenIIIavgprobPLATTgASCElangRust  & \MEQwenIIIavgprobLRgASCElangRust & \bf \MEQwenIIIavgprobLOGRgASCElangRust & \textbf{\MEQwenIIIavgprobIGLBgASCElangRust}      \\ \bottomrule  
\end{tabularx}
\end{table}






\section{Threats to Validity}
We identified the following potential threats to validity in our study.
First, two of our benchmarks, LiveCodeBench and MultiPL-E, contain artificial problems that may not fully reflect real-world software development. In contrast, McEval is based on real-world code samples, which gives confidence that the positive results observed on this benchmark may generalize to practical SE scenarios. 
Second, our analysis focuses on function synthesis and does not cover the full spectrum of code generation tasks. Other software engineering applications may exhibit different calibration behaviors and should be explored in future work.
Finally, a common concern when evaluating LLMs is data leakage. We mitigate this risk to some extent by using two very recent datasets (LiveCodeBench and McEval), for which most samples were published after the models’ knowledge cutoff dates.

\section{Conclusions}
In this paper, we argue that calibrated confidence scores can enable more reliable and developer-friendly integration of LLM-generated code into software development workflows. Building on prior findings that LLMs are often poorly calibrated for code synthesis, we replicated this result for current reasoning models and explored a range of post-hoc calibration methods to address the issue. In particular, we introduce the concept of \emph{multicalibration}, which -- to our knowledge -- has not been previously explored in the context of code generation, and demonstrated its effectiveness compared to classical calibration techniques. 
Our experiments highlight that incorporating structured, code-related group information can significantly improve calibration quality. While our study is not exhaustive, it provides a methodological and empirical foundation for further work on confidence calibration in SE. Future directions include exploring alternative initial scoring functions (e.g., based on distributional features of token likelihoods or on internal LLM states) and extending calibration analysis to broader tasks, such as repository-level code generation or program repair. To support this, we make code and data publicly available as a replication package~\cite{campos2025replication}.

\bibliographystyle{ACM-Reference-Format}

\bibliography{calibration}

\end{document}